\documentclass[a4paper,english,fleqn]{article}
\pdfoutput=1

\usepackage{amsmath}
\usepackage{amssymb}

\makeatletter

\pdfpageheight\paperheight
\pdfpagewidth\paperwidth

\usepackage[english]{babel}
\usepackage{authblk}
\usepackage{amsfonts}
\usepackage{amsthm}
\usepackage{color}
\usepackage{empheq}
\usepackage{graphicx}
\usepackage{float}
\usepackage{bbm}

\usepackage[colorlinks,linkcolor=blue,anchorcolor=black,citecolor=red]{hyperref}

\def\be{\begin{equation}}
\def\ee{\end{equation}}
\def\bc{\begin{center}}
\def\ec{\end{center}}

\newcommand{\N}{{\mathbb{N}}}
\newcommand{\R}{{\mathbb{R}}}
\newcommand{\E}{{\mathbb{E}}}
\renewcommand{\P}{{\mathbb{P}}}

\makeatother

\usepackage{babel}
\begin{document}

%
%
%
%

\title{Current large deviations for partially asymmetric particle systems on a ring}

\author[a]{Paul Chleboun}
\author[b]{Stefan Grosskinsky}
\author[b,c]{Andrea Pizzoferrato}
\affil[a]{Department of Statistics, University of Oxford}
\affil[b]{Mathematics Institute, University of Warwick}
\affil[c]{The Alan Turing Institute, London}

\renewcommand\Authands{ and }

%
%
%

\maketitle
%
%
%
%
\begin{abstract}
We study large deviations for the current of one-dimensional stochastic particle systems with periodic boundary conditions. Following a recent approach based on an earlier result by Jensen and Varadhan, we compare several candidates for atypical currents to travelling wave density profiles, which correspond to non-entropic weak solutions of the hyperbolic scaling limit of the process. We generalize previous results to partially asymmetric systems and systems with convex as well as concave current-density relations, including zero-range and inclusion processes. We provide predictions for the large deviation rate function covering the full range of current fluctuations using heuristic arguments, and support them by simulation results using cloning algorithms wherever they are computationally accessible. For partially asymmetric zero-range processes we identify an interesting dynamic phase transition between different strategies for atypical currents, which is of a generic nature and expected to apply to a large class of particle systems on a ring.

\end{abstract}

%
%
%
%


%
%
%
%
\section{Introduction}

Large deviations of dynamic observables in bulk-driven lattice gases have been a topic of major recent research interest. As summarized in a recent review \cite{lazar1}, most studies focus on the empirical particle current as one of the most important characteristics of nonequilibrium systems in one dimension. 
To derive the large deviation rate function for additive path functionals such as the current, the associated scaled cumulant generating function can be characterized in terms of the leading eigenvalue of a tilted version of the generator of the process \cite{Chetrite2014}. 
In many cases an ansatz for the corresponding eigenfunction can be derived and such microscopic methods have been applied successfully to various models: 
To the asymmetric simple exclusion process (ASEP) \cite{gorissen2012exact,Bodineau2006} based on Bethe-ansatz type techniques related to exact solvability, and also in combination with the matrix product ansatz \cite{derrida2007non}, and to zero-range processes (ZRP) \cite{Harris2005,Harris2013,Hirschberg2015} based on existence of non-homogeneous factorized stationary distributions. 
These studies focus almost entirely on open boundary conditions with only few exceptions covering the periodic case \cite{popkov2010asep, tsobgni2016large}, where microscopic results are difficult to obtain due to temporal correlations \cite{indiansZRP07}. 
Dynamic large deviations have also been studied successfully from a macroscopic point of view using more generally applicable techniques, most notably macroscopic fluctuation theory (MFT) (see \cite{Bertini2014a} and references therein), which can be understood in terms of empirical flows for Markov chains \cite{3bigsfirst,3bigssecond}. The time evolution of the most likely density profile for a given fluctuation of the current is characterized by a variational principle, which can be hard to solve in general, and explicit expressions have only been obtained for some specific models so far \cite{Bertini2014a,lazar1}. A-priori this approach is limited to symmetric or weakly asymmetric bulk dynamics, but full asymmetry can often be covered in the limit of a diverging weak field.

In general, macroscopic approaches rely on a hydrodynamic description of the process in terms of a conservation law. For  asymmetric models this is a hyperbolic equation with weak solutions that can develop shock discontinuities, and for which additional selection criteria are needed to identify a unique (entropic) solution, such as a fixed positive sign for the entropy production (see e.g.\ \cite{Smoller}). 
Provided the 'correct' thermodynamic entropy functional is used, the negative part of the entropy production provides the large deviation rate function for observing a non-entropic weak solution in the scaling limit \cite{Varadhan2004}. This general idea has been proved rigorously only for the ASEP so far \cite{Jensen,Vilensky2008}. In \cite{DerriBodJV}, this has been applied heuristically to obtain the rate function for lower current deviations for the ASEP, which are realized by phase separated (travelling wave) profiles where two regions of different densities are separated by two shock discontinuities, in agreement with exact microscopic results. In recent work \cite{pizzo} this approach has been shown to apply also for totally asymmetric ZRPs with concave current-density relation, where the validity can be limited by a crossover to condensed profiles in certain models constituting a dynamic phase transition.

The point of this paper is to highlight the general applicability of the approach in \cite{pizzo} for general asymmetric particle systems with stationary product distributions.
This is illustrated by applications to the inclusion process (IP) with convex current-density relation, and ZRPs with a concave relation and partially asymmetric dynamics. 
Based on first results in \cite{drouffe98,evansBrazil}, ZRPs have been studied in detail in the context of a condensation transition in homogeneous systems.
Condensation arises due to particle interactions when the density exceeds a critical value (see e.g. \cite{Evans2005,godreche,godreche2012condensation} and references therein), including also rigorous mathematical results (see e.g. \cite{Chleboun2014} and references therein).
The IP has been introduced more recently in \cite{redig} and was also studied in the context of condensation \cite{grv}, together with certain variations (see \cite{chau}). 
For this paper condensation will only play a minor role, but both models exhibit factorized stationary distributions and constitute paradigmatic particle systems with unbounded local density and convex or concave current-density relations.

We characterize the rate function for upper current deviations for the IP by a variational principle in terms of travelling wave profiles, which can be optimised numerically and agrees well with simulation data. It turns out that condensed profiles do not contribute to large deviation events in the IP. For partially asymmetric dynamics we derive a general relation for the cost functionals of travelling wave and condensed profiles and their totally asymmetric counterparts. We illustrate this result for a condensing ZRP which exhibits the most interesting behaviour in this context. 
As discussed in detail in \cite{pizzo} (see also Figure \ref{fig:illu} in Section \ref{sec:2}), for systems with concave (convex) current-density relation only lower (upper) current deviations are accessible via phase separated profiles. Outside this range, deviations of the current are usually associated to hyperuniform states with long-range correlations \cite{jack2015hyperuniformity, karevski2016conformal}. We discuss such candidates for inclusion and partially asymmetric ZRPs, covering the full range of current deviations, and identifying interesting transitions between different types of optimal states.  This leads to a complete characterization of the rate function for both models, and we discuss how the generic nature of our approach can be applied in general particle systems.

The paper is structured as follows. In Section \ref{sec:2}, we set up notation for stochastic lattice gases, their stationary distributions and large deviations for the empirical current. In Section \ref{sec:inclusion} we apply the approach developed in \cite{pizzo} to the upper current deviations in the IP, and also discuss profiles for lower deviations. In Section \ref{sec:PA} we extend the approach to partially asymmetric systems, illustrate it for a particular ZRP, and derive a generic form of the rate function for deviations outside the range of phase separated profiles. We end by discussing the generic nature and applicability of our approach in Section \ref{sec:discussion}.

%
%
%
%
\section{Definitions and Setting}\label{sec:2}
%
%
\subsection{Stochastic particle systems on a ring}
Consider a one-dimensional lattice $\Lambda$ with $\left|\Lambda\right| =L\in\mathbb{N}$ sites and  periodic boundary conditions. Each site $x\in\Lambda$ can carry an integer number of particles $\eta_x\in\mathbb{N}_0$, 
and a configuration of the system is denoted by $\eta=\left(\eta_{1},\eta_{2},...,\eta_{L}\right)\in X_L$, where $X_L =\mathbb{N}_0^{\Lambda}$ is the configuration space. 
We consider processes with nearest-neighbour particle jumps from sites $x$ to $y=x\pm 1$ at rate $p(x,y) u(\eta_x)v(\eta_y)$, focusing on translation invariant dynamics with $p(x,y)=p\delta_{y,x+1} +q\delta_{y,x-1}$ (we consider periodic boundary conditions so addition is taken modulo $L$).
The dynamics of the process can be described by the generator
\begin{equation}\label{gen}
	\mathcal{L}f\left(\eta\right)=\sum_{x,y\in\Lambda}p(x,y)u\left(\eta_{x}\right)v\left(\eta_{y}\right)\left[f\left(\eta^{x,y}\right)-f\left(\eta\right)\right] \ ,
\end{equation}
for test functions $f:X_L \to \mathbb{R}$. 
As usual, we denote by $\eta^{x,y}$ the configuration obtained from $\eta$ after a particle jumps from site $x$ to $y$, i.e. $\eta^{x,y}_z =\eta_z -\delta_{z,x} +\delta_{z,y}$. 
Since we consider only finite lattices there are no restrictions on the observable $f$, see e.g.\ \cite{Andjel1982,balazs} for details on infinite lattices for particular models. 
To avoid degeneracies and for later convenience we assume that the rates are in fact defined by smooth functions $u,v:\R\to [0,\infty )$ with
\begin{equation}\label{eq:transrule}
u\left(n\right)=0\mbox{ if and only if }n=0\quad\mbox{and}\quad v\left(n\right)>0\mbox{ for all }n\geq 0\ .
\end{equation}
The total number of particles is a conserved quantity under the dynamics, and the process is irreducible on the state space $X_{L,N}\coloneqq\left\{ \eta\in X_L:\sum_{x\in\Lambda}\eta_{x}=N\right\}$ for each fixed $N\geq 0$.
We denote the process by $(\eta (t) :t\geq 0)$, with the usual notation for the path space distribution $\P$ and the corresponding expectation $\E$.

While our main results are applicable more generally (as discussed in Section \ref{sec:discussion}), we focus the presentation on two particular models, namely zero-range processes (ZRP), where
\begin{equation}
u(n)\mbox{ is arbitrary, and }v(n)\equiv 1\ ,
\label{zrprate}
\end{equation}
and the inclusion process (IP), where for some parameter $d>0$
\begin{equation}
u(n)=n\quad\mbox{and}\quad v(n)=(d+n)\ .
\label{iprate}
\end{equation}

It is well known (see e.g.\ \cite{Chleboun2014} and references therein) that both 
models admit stationary product distributions, the so-called grand-canonical measures, 
\begin{equation}\label{pm}
	\nu_{\phi}^L\left[d\eta\right]\coloneqq\prod_{x\in\Lambda}\nu_{\phi}\left(\eta_x\right)d\eta
\end{equation}
with a fugacity parameter $\phi\geq 0$ controlling the particle density. 
The mass function of the single site marginal with respect to the counting measure $d\eta$ on $X_L$,  is given by
\begin{equation}
\nu_{\phi} \left(\eta_{x}\right)=\frac{1}{z\left(\phi\right)}w\left(\eta_{x}\right)\phi^{\eta_{x}}\ ,
\label{marginals}
\end{equation}
with stationary weights
\begin{equation}\label{weights}
	w\left(\eta_{x}\right)=\prod_{k=1}^{\eta_{x}}\frac{v(k-1)}{u(k)} \quad\text{where}\quad w\left(0\right)=1\ ,
\end{equation}
and normalization given by the (grand-canonical) partition function
\begin{equation}\label{eq:partfct}
	z\left(\phi\right)=\sum_{n=0}^{\infty}w\left(n\right)\phi^{n} \ .
\end{equation}
The distributions $\nu_\phi$ exist for all $\phi\geq 0$ such that $z(\phi )<\infty$, and we denote by $\phi_c \in (0,\infty ]$ the radius of convergence of $z(\phi )$, which we assume to be strictly positive. A convenient sufficient condition to ensure this for ZRPs is that the jump rates are asymptotically bounded away from $0$, i.e. $\liminf_{k\to\infty} u(k)>0$ (see e.g.\ \cite{Chleboun2014}).

Under the grand-canonical measures the total particle number is random, and the fugacity parameter controls the average density
\begin{equation}\label{density}
	R\left(\phi\right)\coloneqq\left\langle\eta_{x}\right\rangle_\phi \coloneqq \sum_{n\in\N} \nu_\phi (n)n =\phi\,\partial_{\phi}\ln z\left(\phi\right)\ ,
\end{equation}
where we use the notation $\langle\cdot\rangle_\phi$ for expectations w.r.t.\ the distribution $\nu_\phi$. 
In general, $\ln z\left(\phi\right )$ is known to be a log-convex function with
\[
(\phi\,\partial_\phi )^2\ln z(\phi )=\phi\,\partial_\phi R(\phi )=\langle \eta_x^2 \rangle_\phi -\langle \eta_x \rangle_\phi^2 > 0\quad\mbox{for all }\phi >0\ ,
\]
and $R\left(\phi\right )$ is striclty increasing in $\phi$ and continuous, with $R\left(0\right) =0$ and largest value
\begin{equation}
\rho_c :=\lim_{\phi\nearrow\phi_c} R(\phi )\in (0,\infty ]\ .
\label{rhocdef}
\end{equation}
This is also called the critical density, and if finite, the system only has homogeneous stationary product measures for a bounded range of densities $[0,\rho_c ]$, with $\nu_{\phi_c}$ being the maximal invariant measure. 
We denote by
\begin{equation}\label{phirho}
\Phi\left(\rho\right)\quad\mbox{the inverse of}\quad R(\phi )\ ,
\end{equation}
which will be made use of later.

For the IP above quantities can be computed explicitly (see e.g.\ \cite{gredvaf11}), and for any $d>0$ we have
\begin{equation}\label{ippart}
z(\phi)=(1-\phi)^{-d}\ ,\quad R(\phi )=\frac{d\phi }{1-\phi}\quad\mbox{with}\quad \phi_c =1\ ,
\end{equation}
and thus $\rho_c =\infty$. For ZRPs $\rho_c <\infty$ is possible for particular choices of rates $u(n)$, as is discussed in Section \ref{sec:zrp}.

On the finite state space $X_{L,N}$ the process is irreducible, and the corresponding unique stationary distributions are the canonical stationary measures. 
They can be expressed by conditioning the grand-canonical distribution to a fixed number of particles as
\begin{equation}\label{canmeas}
\pi_{L,N}\left[d\eta\right]\coloneqq\nu_{\phi}^{L}\left[d\eta\left|X_{L,N}\right.\right] =\frac{\mathbbm{1}_{X_{L,N}}(\eta)}{Z_{L,N}}\prod_{x\in\Lambda_{L}}w\left(\eta_x \right)\, d\eta\ .
\end{equation}
Here $Z_{L,N}\coloneqq\sum_{\eta\in X_{L,N}}\prod_{x} w\left(\eta_x\right)$ is the canonical partition function and we note that \eqref{canmeas} is independent of $\phi$. 
We denote  the expectation w.r.t.\ $\pi_{L,N}$ by $\langle \,\cdot\, \rangle_{L,N}$. Note that the conditioning on a fixed total number of particles introduces (weak) negative correlations between occupation numbers, which are independent of spatial distances, i.e.
\begin{equation}\label{corr}
\langle \eta_x \eta_y \rangle_{L,N} -\Big(\frac{N}{L}\Big)^2 \equiv c_{L,N} <0\ ,\quad\mbox{independently of }x\neq y\in\Lambda\ .
\end{equation}
While canonical measures exist for general conservative particle systems, \eqref{canmeas} and \eqref{corr} only hold for systems that exhibit grand-canonical product measures, which we concentrate on in this paper.

%
%
\subsection{Current large deviations}

In our setting of translation-invariant nearest-neighbour dynamics the average stationary current w.r.t. to the canonical measure is defined as
\begin{equation}\label{cancurr}
J_{L,N} \coloneqq (p-q)\big\langle u(\eta_x)v(\eta_{x+1})\big\rangle_{L,N} \ ,
\end{equation}
where we have used that spatial correlations are site independent \eqref{corr}. 
Under the grand-canonical measures we have for ZRPs
\begin{equation}\label{current}
	J\left(\rho\right)\coloneqq (p-q)\left\langle u(\eta_x )\right\rangle_{\Phi (\rho )}=(p-q)\Phi\left(\rho\right)\ ,
\end{equation}
with $\Phi\left(\rho\right)$ given in (\ref{phirho}), as a direct consequence of the form of the stationary weights (\ref{weights}). For IPs we simply get the explicit expression
\begin{equation}\label{current_ip}
J\left(\rho\right)\coloneqq(p-q)\left\langle u(\eta_x)v(\eta_{x+1})\right\rangle_{\Phi (\rho )}=(p-q)\rho (d+\rho )\ .
\end{equation}
Due to the equivalence of ensembles (see e.g. \cite{Chleboun2014} and references therein), both stationary currents are equivalent in the thermodynamic limit, i.e.\ for all $\rho <\rho_c$
\begin{equation}
J_{L,N} \to J(\rho )\quad\mbox{as }L,N\to\infty\quad\mbox{with }N/L\to\rho\ .
\label{equivalence}
\end{equation}
The (random) empirical current averaged over sites up to time $t>0$ is given by
\begin{equation}
\mathcal{J}^{L}\left(t\right)\coloneqq\frac{1}{L}\sum_{x}\mathcal{J}_{x,x+1}^{L}\left(t\right)
\label{emcu}
\end{equation}
where the net current across the bond $x,x+1$ per unit time is given by
\begin{equation}
\mathcal{J}_{x,x+1}^{L}\left(t\right)\coloneqq
\frac{1}{t}\sum_{s\in [0,t]}\left(\eta_{x}(s)-\eta_{x}(s^-)\right)^2 \left(\eta_{x+1}(s)-\eta_{x+1}(s^-)\right)\ .
\end{equation}
Note that $\mathbb{P}$-almost surely this sum has only finitely many non-zero terms which are $\pm 1$ depending on the direction of the particle jump.

For fixed $L$ and $N$ the stochastic particle systems are finite-state, irreducible Markov chains on $X_{L,N}$, and a general approach in \cite{3bigsfirst, 3bigssecond} implies a large deviation principle (LDP) for the empirical current (\ref{emcu}) in the limit $t\to\infty$ (see also \cite{pizzo} for more details.). 
We denote the associated rate function by $I^L :\R\to [0,\infty ]$, and for all regular intervals $A\subseteq \R$ we have 
(see e.g.\ \cite{touchette2017}) 
\begin{equation}\label{ldp}
\frac{1}{t}\log\mathbb{P}\left[\mathcal{J}^{L}\left(t\right)\in A\right]\to \inf_{j\in A} I^L (j) \quad\mbox{as }t\to\infty\ .
\end{equation}
Informally, one often writes \eqref{ldp} as
\[
\mathbb{P}\left[\mathcal{J}^{L}\left(t\right)\approx j\right]\asymp e^{-tI^L (j)}\ .
\]
The approach in \cite{3bigsfirst, 3bigssecond} based on the contraction principle by a linear mapping also implies that the rate function $I^L (j)$ is in fact convex. 
Generalizing recent results for totally asymmetric ZRPs \cite{pizzo} and previous work on exclusion \cite{BodADD,Bodineau2006,Jensen,Vilensky2008}, our main result is a derivation of the rate function for diverging system size
\begin{equation}
I (j)=\lim_{L\to\infty} I^L (j)\ ,
\label{ratefu}
\end{equation}
including lower and upper deviations of the current under partial asymmetry. 
While for IPs the current \eqref{current_ip} is a strictly convex function of $\rho$, for ZRPs we focus on examples where $J(\rho)$ \eqref{current} is concave and non-linear. 

In addition to macroscopic arguments based on the Jensen-Varadhan approach and heuristics for particle profiles, we also present simulation results from cloning algorithms based on the grand-canonical or tilted path ensemble 
\cite{Giardina2011, Harris2013, Chetrite2014}. These provide access to the scaled cumulant generating function defined as
\begin{equation}
\lambda^L\left(k\right)\coloneqq\lim_{t\to\infty}\frac{1}{t}\ln \mathbb{E}\left[e^{tk\mathcal{J}^{L}\left(t\right)}\right] ,
\label{eq:momgen}
\end{equation}
and since the rate function $I^L$ is convex, it is then given by the Legendre-Fenchel transform
\begin{equation}
\label{eq:leg}
I^{L}\left(j\right)=\sup_{k\in\mathbb{R}}\left\{ kj-\lambda^L\left(k\right)\right\} \ .
\end{equation}

In Section \ref{sec:inclusion} we study the totally asymmetric IP with $p=1-q=1$. In analogy to \cite{pizzo}, we will see that due to the convex current $J(\rho)$ upper large deviations for large $L$ are dominated by phase separated states which are non-entropic weak solutions of the hydrodynamic limit equation. 
Our second generalization of \cite{pizzo} concerns partially asymmetric dynamics ($1/2<p<1$) of ZRPs, where we establish a full picture for a condensing example including conditioning on negative currents against the bias. 

%
%
%
%
\subsection{Phase separated profiles for current large deviations\label{sec:twprofiles}}

It is well established that the large-scale dynamics of asymmetric particle systems of the form \eqref{gen} in hyperbolic scaling $y=x/L,\ \tau =t/L$ in the hydrodynamic limit is described by the conservation law for the density field $\rho\left(y,\tau\right)=\lim_{L\to\infty}\mathbb{E}\left[\eta_{yL}\left(\tau L\right)\right]$,
\begin{equation}\label{pde}
		\frac{\partial}{\partial \tau}\rho \left(y,\tau\right)+\frac{\partial}{\partial y}J\left(\rho\left(y,\tau\right)\right)=0\quad y\in\mathbb{T},\ \tau\geq 0\ .
\end{equation}
Here $\mathbb{T}$ denotes the unit torus corresponding to periodic boundary conditions. 
This has been proved rigorously for attractive processes which preserve stochastic order in time using coupling techniques (see e.g. \cite{Landim,saada} and references therein). Models of the form \eqref{gen} are attractive if and only if $u$ is a non-decreasing and $v$ is a non-increasing function of the number of particles. Note that the condition on $v$ does not hold for IPs, while ZRPs are attractive whenever $u$ is non-decreasing. In the absense of attractivity there are only partial results for ZRPs with sublinear (and possibly decreasing) jump rates using relative entropy methods (see e.g.\ \cite{Landim}, Chapter 5), 
and partial results on symmetric counterparts for ZRP \cite{stamatakis} and IPs \cite{opoku}. 
Still, the scaling limit (\ref{pde}) is believed to hold also for asymmetric systems of the form \eqref{gen} under more general conditions, see e.g.\ \cite{Schutz2007} for heuristic results on condensing ZRPs. 

Solutions to \eqref{pde} can develop shock discontinuities even for smooth initial data, which leads to the concept of weak solutions which satisfy an integrated version (see e.g.\ \cite{Smoller}, Section 15). 
These are in general not unique, and the physically relevant ones
may be selected by the entropy condition developed by Kruzkov (see e.g.\ \cite{laxbook}). 
Consider a regular convex function $h\left( \rho\right)$, called entropy, with corresponding {entropy flux $g\left( \rho\right)$ such that
\begin{equation}
g'\left(\rho\right)=J'\left(\rho\right)h'\left(\rho\right).
 \label{eflux}
 \end{equation}
Weak solutions are called entropy solutions if for all entropy-entropy flux pairs
\begin{equation}
		\frac{\partial}{\partial \tau}h(\rho \left(y,\tau\right) )+\frac{\partial}{\partial y}g\left(\rho\left(y,\tau\right)\right)\geq 0\ ,
\label{kruzkov}
\end{equation}
in a weak sense, and such solutions are uniquely determined by their initial data. Note that for smooth solutions equality holds in (\ref{kruzkov}) and due to \eqref{eflux} entropy is conserved. 
Entropy is not conserved across shocks, when the solution jumps from a value $\rho_l$ on the left to $\rho_r \neq\rho_l$ on the right. By conservation of mass it is easy to show that such a shock travels with velocity
\begin{equation}
v_{s} (\rho_l ,\rho_r) =\frac{J\left(\rho_r\right)-J\left(\rho_l\right)}{\rho_r-\rho_l}\ .
\label{speed}
\end{equation}
Shocks that are stable under the time evolution for entropy solutions fulfill
\begin{equation}
J'(\rho_l )>v_s (\rho_l ,\rho_r )>J'(\rho_r )\ .
\end{equation}
Unstable shocks turn into so-called rarefaction fans, which are travelling wave solutions interpolating continuously between the two densities $\rho_l$ and $\rho_r$ (see \cite{Smoller} for details). So for convex $J(\rho)$, $J'(\rho)$ is monotone increasing and down shocks ($\rho_l >\rho_r$) are stable, as is the case in the IP. Analogously, for concave $J(\rho)$ up shocks are stable. 
The entropy production rate across a shock with $\rho_l \neq\rho_r$ and speed $v_s$ (\ref{speed}) is given by integrating (\ref{kruzkov}) over space,
\begin{equation}
\mathcal{F}\left(\rho_l ,\rho_r \right) \coloneqq g\left(\rho_l \right)-g\left(\rho_r \right)-\frac{J\left(\rho_r\right)-J\left(\rho_l\right)}{\rho_r-\rho_l}\left(h\left(\rho_l\right)-h\left(\rho_r\right)\right)\ .
\label{entprod}
\end{equation}
Note that this rate would be negative across unstable shocks. 

For the asymmetric simple exclusion process (ASEP) it was shown in \cite{Jensen} and \cite{Vilensky2008} that the large deviation rate function to observe a non-entropic weak solution over a fixed macroscopic time interval $[0,\tau ]$ in the limit $L\to\infty$ is given by the accumulated negative part of the entropy production \eqref{entprod}, choosing $h$ to be the thermodynamic entropy of the system. \cite{Varadhan2004}. 
For stochastic particle systems with stationary product measures of the form \eqref{pm} 
the thermodynamic entropy is given by the Legendre transform of the free energy,
\begin{equation}\label{eq:entrflux}
h\left(\rho\right)=\rho\ln\Phi\left(\rho\right)-\ln z\left(\Phi\left(\rho\right)\right)\ .
\end{equation}
This result has been applied heuristically in \cite{Bodineau2006} for ASEP, and recently in \cite{pizzo} for totally asymmetric ZRPs with concave $J(\rho)$, to derive the large deviation rate function for current fluctuations \eqref{ratefu}. For fixed, large system size $L$, lower current deviations on a ring are realized by phase separated travelling wave step profiles with two densities $\rho_1 <\rho_2$, as illustrated in Fig.\ \ref{fig:illu} (top). 
The probabilistic cost to realize such a profile does not depend on system size since only the non-entropic shock has to be stabilized, which is a localized object. This cost is equal to the entropy production across the reversed stable shock given by $\mathcal{F} (\rho_1 ,\rho_2 )$, which is equal to $-\mathcal{F} (\rho_2 ,\rho_1 )$ by obvious symmetry in \eqref{entprod}.

\begin{figure}[t!]
	\begin{center}
	\includegraphics[width=0.55\textwidth]{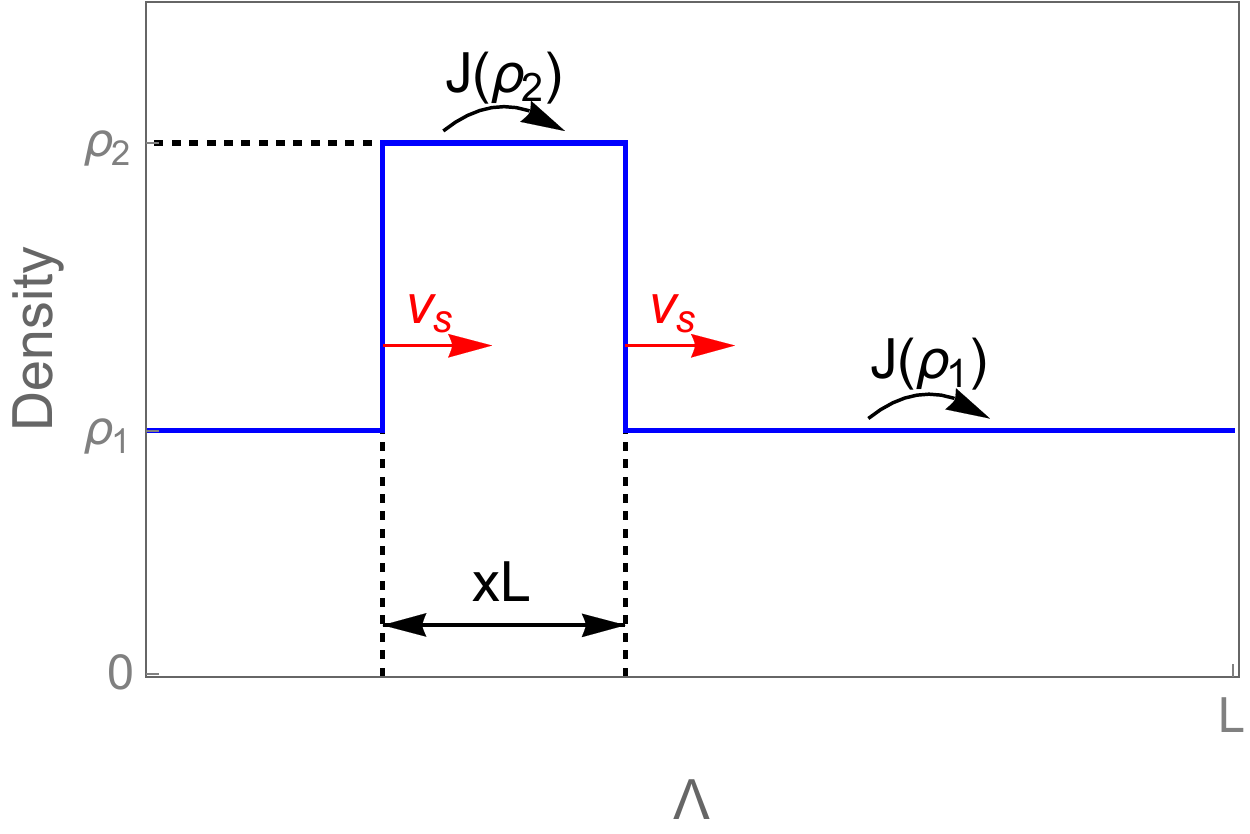}\\	\mbox{\includegraphics[width=0.47\textwidth]{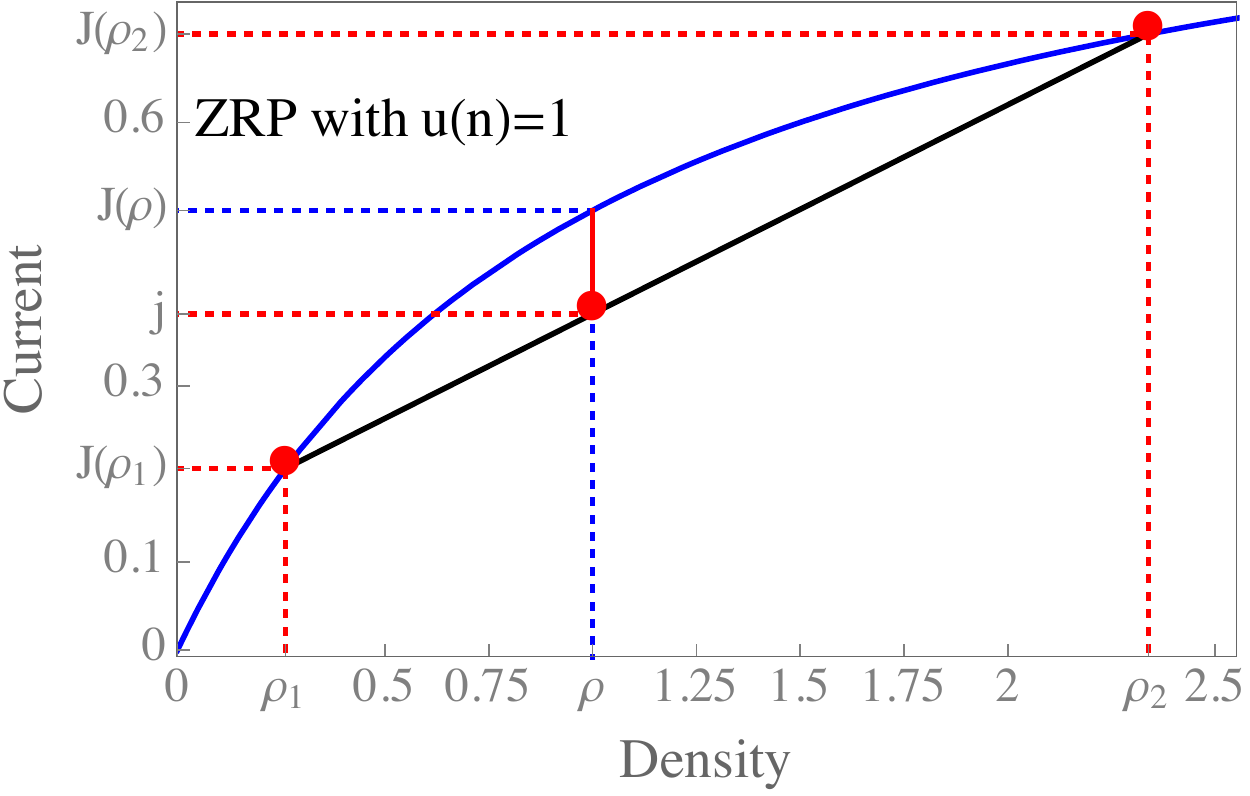}\quad\includegraphics[width=0.47\textwidth]{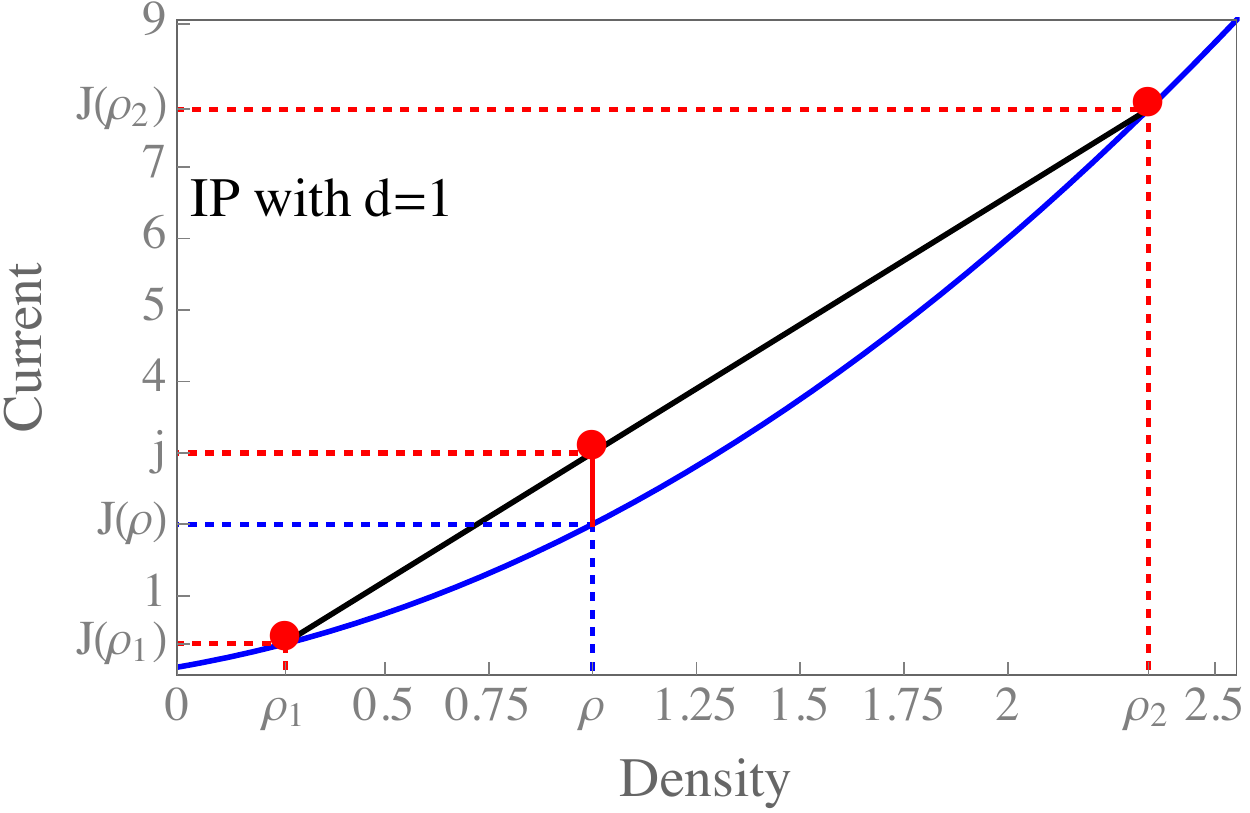}
		}
	\end{center}
	\caption{\label{fig:illu}
		The top illustrates a travelling wave profile with densities $\rho_1 <\rho <\rho_2$, travelling at speed $v_s$ \eqref{speed} to the right. 
		The bottom two plots illustrate the consistency relations \eqref{eq:sistIPdenscur} for $j<J(\rho)$ for concave (left) and $j>J(\rho)$ for convex stationary current. The blue curves $J(\rho)$ correspond to a ZRP with constant rates $u(n)=1$, $n\geq 1$ (left), and an IP with $d=1$ (right). The grey secant intersects $J(\rho)$ at $\rho_1$ an $\rho_2$ for an exemplary admissible pair of densities.
	}
\end{figure}

%
%
%
%


Travelling wave profiles consist of two phase separated regions with densities $\rho_1 <\rho_2$. Denoting by $x\in [0,1]$ the volume fraction of the high density phase, such a profile has typical current $j$ given by
\begin{alignat}{2}\label{eq:sistIPdenscur}
j=&\left(1-x\right)J\left(\rho_{1}\right)+xJ\left(\rho_{2}\right)\nonumber\\
\rho=&\left(1-x\right)\rho_{1}+x\rho_{2},
\end{alignat}
as illustrated in Fig.\ \ref{fig:illu} (bottom), where $\rho$ denotes the average density associated to the profile. 
By eliminating the variable
\begin{equation}\label{eq:hdensfrac}
x=\frac{j-J(\rho_1)}{J(\rho_2)-J(\rho_1)},
\end{equation}
the constraints \eqref{eq:sistIPdenscur} can be re-written as
\begin{equation}
G\left(\rho_{1},\rho_{2}\right)\coloneqq\frac{\rho\left(J(\rho_2)-J(\rho_{1})\right)-J(\rho_2) \rho_1 +J(\rho_1 )\rho_2}{\rho_2 -\rho_1}=j\ .
\label{cond}
\end{equation}
This implicitly defines a one-parameter family of admissible densities $(\rho_1 ,\rho_2 )$ for travelling wave profiles with given $\rho$ and $j\neq J(\rho)$, which can often be solved explicitly in paricular cases such as the IPs (see Section \ref{sec:inclusion}). We consider models where the current-density relation $J(\rho)$ is convex or concave over the whole density region, and it is clear from Fig.\ \ref{fig:illu} that admissible densities only exist if $j>J(\rho)$ or $j<J(\rho)$, respectively. 

Due to this convexity (concavity) assumption, one of the shocks in a travelling wave profile is stable, and the large deviation cost of the profile is given by the negative entropy production across the non-entropic shock. Due to symmetry of the functional \eqref{entprod} we can write this in general as $|\mathcal{F}(\rho_1 ,\rho_2)|$. The associated rate function for fixed total density $\rho$ is then given by minimizing \eqref{entprod} subject to the constraint \eqref{cond} over all possible density pairs $\rho_1 \leq\rho\leq\rho_2$, that is 
\begin{equation}
\label{twcost}
E_{tw} (j)\coloneqq\inf_{\rho_1\leq\rho\leq\rho_2}\big\{ |\mathcal{F}(\rho_1 ,\rho_2)|\, :\, G\left(\rho_{1},\rho_{2}\right)=j\big\}\in [0,\infty ]\ .
\end{equation}
Depending on the specifics of $\mathcal{F}$ and $G$ in a given example, the minimizer in \eqref{twcost} is often unique and in the interior of the density domain, and can be found using standard Lagrange multipliers as we will see later. But the global minimum can also be attained at the boundary, as is the case for certain condensing systems as discussed in Section \ref{sec:zrp} and in more detail in \cite{pizzo}. 
If for a given $j$ the condition \eqref{cond} cannot be fulfilled the minimization in \eqref{twcost} leads to $E_{tw} (j) =\infty$, and such a current deviation cannot be realized with a travelling wave profile.

There are of course many other strategies to realize large deviations for the empirical current. Often these involve a global change of the dynamics leading to costs proportional to the system size, and these are only relevant whenever travelling wave profiles are not accessible and discussed for particular cases in later sections. One particular strategy also of a local nature are condensed profiles, which can have costs independent of the system size in systems with bounded rates, as discussed in more detail in Section \ref{sec:zrp} for ZRPs.
For systems with either convex or concave $J(\rho )$, travelling wave profiles with more than one up and down shock are more costly than the simple one shown in Figure \ref{fig:illu} (top), and do not contribute to typical large deviation events.

\section{Totally asymmetric inclusion process\label{sec:inclusion}}

In this section we follow the same arguments presented in \cite{pizzo} to derive the minimal cost of travelling wave profiles and the rate function for IPs \eqref{iprate} under total asymmetry, i.e.\ $p=1-q=1$.

\subsection{Upper current deviations via travelling waves}

The main difference to previous results for ZRPs with concave $J(\rho)$ is that for the IP $J\left(\rho\right) =\rho (d+\rho)$ \eqref{current_ip} is convex. So down shocks are entropic and stable, and with \eqref{speed} have velocity
\begin{equation}\label{eq:vIP}
v_{s}\left(\rho_{1},\rho_{2}\right)=d+\rho_{1}+\rho_{2}.
\end{equation}
As can be seen from Figure \ref{fig:illu} (right), all values $j>J(\rho)$ are accessible by travelling wave profiles, whereas $j<J(\rho)$ are not due to convexity of $J(\rho)$. 
For given $\rho$ and $j>J(\rho )$ we explicitly solve \eqref{cond} which takes the form
\begin{equation}\label{eq:GIP}
G\left(\rho_{1},\rho_{2}\right)=d\rho-\rho_{1}\rho_{2}+\rho\left(\rho_{1}+\rho_{2}\right)=j,
\end{equation}
to get the monotone increasing relationship
\begin{equation}\label{eq:rho2fromG}
\rho_{2}\left(\rho_{1}\right)=\frac{j-\rho\left(d+\rho_{1}\right)}{\rho-\rho_{1}}\to \left\{
\begin{array}{cl}
(j-d\rho )/\rho &,\ \mbox{as }\rho_1 \to 0\\
\infty &,\ \mbox{as }\rho_1 \to\rho
\end{array}
\right.\ .
\end{equation}

To determine the exponential cost in the minimization problem \eqref{twcost}, we first find an explicit expression for the entropy production \eqref{entprod}. We find the thermodynamic entropy \eqref{eq:entrflux} using explicit expressions summarized in \eqref{ippart},
\begin{equation}
h\left(\rho\right)=\rho\ln\left(\frac{\rho}{d+\rho}\right)+d\ln\left(\frac{d}{d+\rho}\right)\ ,
\end{equation}
and the corresponding flux from \eqref{eflux}
\begin{equation}
g\left(\rho\right)=\rho\left[\left(d+\rho\right)\ln\left(\frac{\rho}{d+\rho}\right)-d\right] .
\end{equation}
So the Jensen-Varadhan rate function \eqref{entprod} for the IP is given by
\begin{equation}\label{eq:JVforIP}
\mathcal{F}\left(\rho_{1},\rho_{2}\right)=d\left(d{+}\rho_{1}{+}\rho_{2}\right)\ln\left(\frac{d+\rho_{1}}{d+\rho_{2}}\right)+\rho_{1}\rho_{2}\ln\left(\frac{\rho_{2}(d+\rho_{1})}{\rho_{1}(d+\rho_{2})}\right)+d\left(\rho_{2}{-}\rho_{1}\right).
\end{equation}
Since down shocks are stable due to convexity of $J(\rho)$, we have that $\mathcal{F}(\rho_2 ,\rho_1)>0$ and the travelling wave rate function \eqref{twcost} is given by 
\begin{equation}\label{etwip}
E_{tw}\left(j\right)=\inf_{\rho_1 \leq\rho\leq\rho_2}\left\{ \mathcal{F}\left(\rho_{2},\rho_{1}\right):G\left(\rho_{1},\rho_{2}\right)=j\right\} \in [0,\infty )
\end{equation}
for all fixed $\rho >0$ and $j>J(\rho)$. 
The minimizer $(\rho_1^o ,\rho_2^o )$ of the preceding expression can be obtained from the standard system of equations for local minimization under constraints
\begin{equation}\label{eq:optIPprobl}
\left\{ \begin{array}{c}
\partial_{1}\mathcal{F}\left(\rho_{2},\rho_{1}\right)\partial_{2}G\left(\rho_{1},\rho_{2}\right)-\partial_{2}\mathcal{F}\left(\rho_{2},\rho_{1}\right)\partial_{1}G\left(\rho_{1},\rho_{2}\right)=0\\
G\left(\rho_{1},\rho_{2}\right)=j
\end{array}\right. ,
\end{equation}
where the first equation can be written explicitly as
\begin{equation}
\left(\rho_{1}-\rho_{2}\right)\left[\rho\ln\left(\frac{\rho_{2}}{d+\rho_{2}}\frac{d+\rho_{1}}{\rho_{1}}\right)+d\ln\left(\frac{d+\rho_{1}}{d+\rho_{2}}\right)\right]=0.
\end{equation}
We notice that of course for $\rho_1=\rho_2 =\rho$ this minimization condition is satisfied, which corresponds to the stationary regime with $j=J\left(\rho\right)$. The solution $\left(\rho_1^o,\rho_2^o\right)$ of the optimization problem in \eqref{eq:optIPprobl} for general $j>J(\rho )$ is unique and can be obtained numerically, as is illustrated in Figure \ref{fig:IPoptpathANDxv} together with the plots of the optimal shock speed $v_s\left(\rho_1^o,\rho_2^o\right)$ and volume fraction of the high density phase \eqref{eq:hdensfrac}.
\begin{figure}[t]
	\centering
	\mbox{\includegraphics[width=0.48\textwidth]{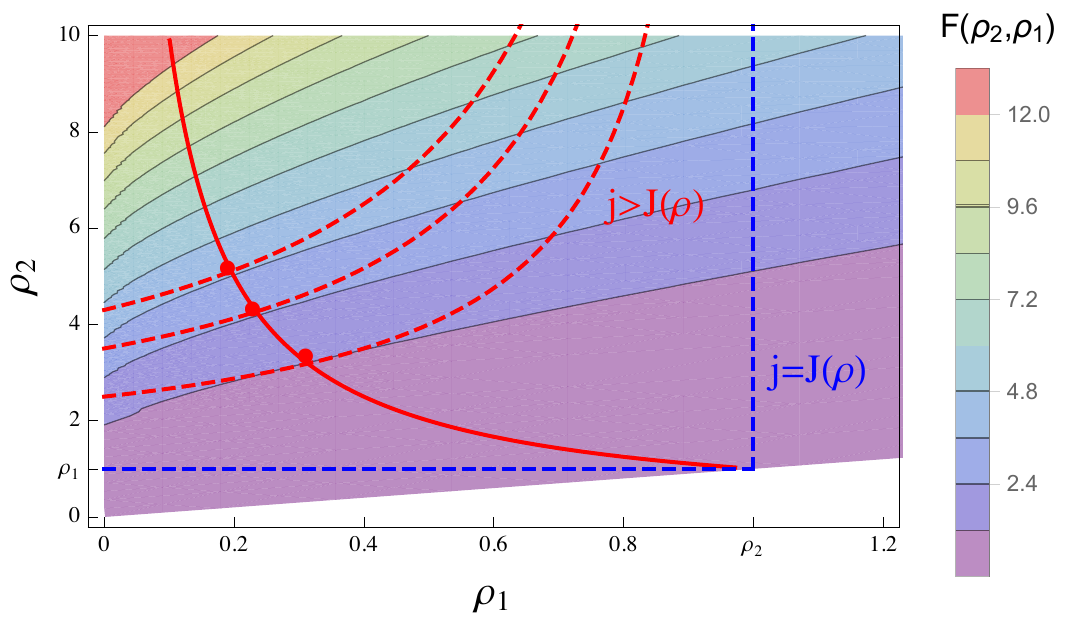}\quad\includegraphics[width=0.48\textwidth]{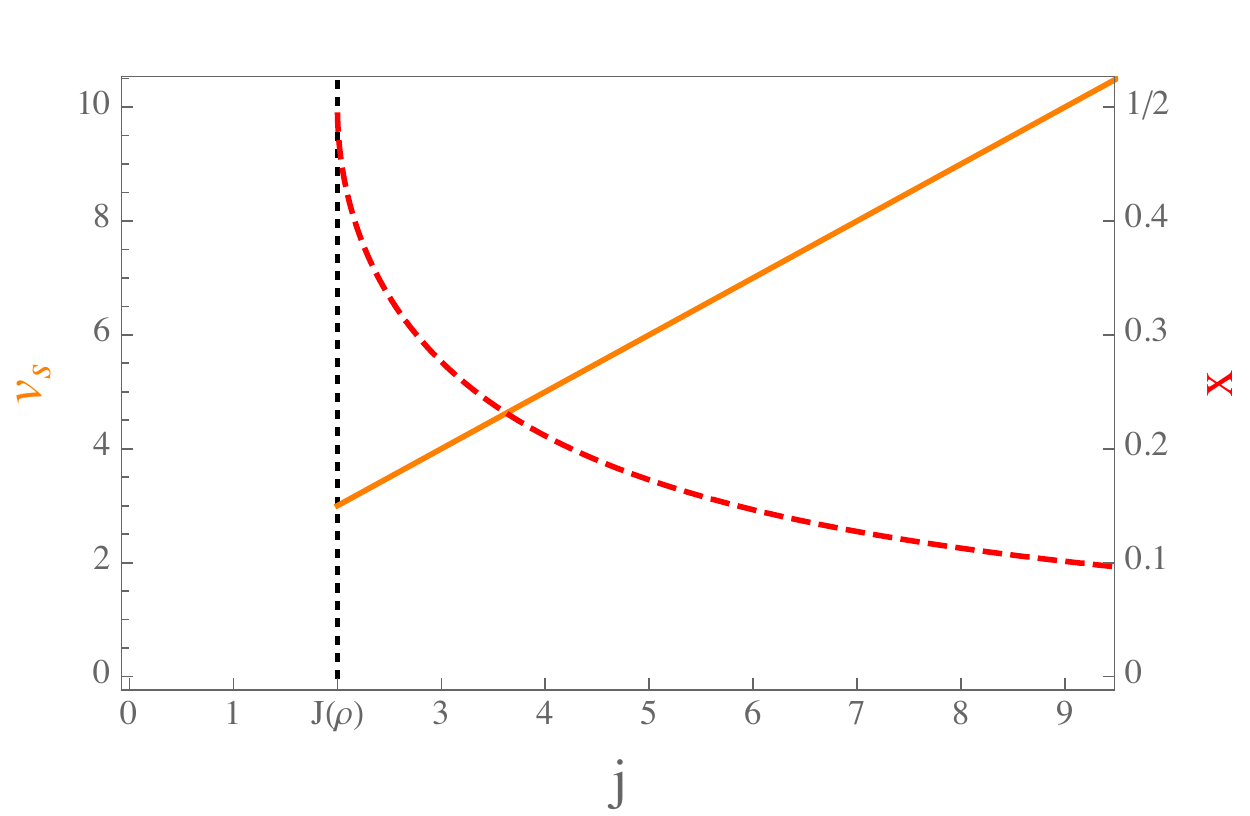}}
	\caption{Optimal cost for an IP with $\rho=1$ and $d=1$. (Left) The contour plot of the cost function \eqref{eq:JVforIP} is in the background, red dotted constrained curves are given by \eqref{eq:rho2fromG} for different values of $j$. The full red curve consists of the optimal pair of points $\left(\rho_1^o,\rho_2^o\right)$ for different values of $j$. (Right) Given the optimal density pairs we plot the profile speed $v_s$ (orange curve) from \eqref{eq:vIP} and the volume fraction $x$ (red dotted curve) from \eqref{eq:hdensfrac}. For $j=J(\rho )$ the system is equally split in the high and low density phase with $x=1/2$, and $x$ decreases with increasing $j$.}\label{fig:IPoptpathANDxv}
\end{figure}

Substituting the solution $\left(\rho_1^o,\rho_2^o\right)$ in the JV function $\mathcal{F}\left(\rho_2^o,\rho_1^o\right)$ for values $j>J(\rho )$ gives rise to a monotone increasing cost function $E_{tw}(j)$, illustrated in Figure \ref{fig:costandSCGFip}, together with its associated SCGF and comparison to simulation data. 
As we discuss below, all conceivable candidates for lower current deviations have associated costs proportional to the system size $L$. Therefore, our first main result is that the large deviation rate function \eqref{ratefu} in the limit $L\to\infty$ is given by 
\begin{equation}
I(j)=\left\{\begin{array}{cl}
E_{tw}(j) &,\ \mbox{for }j\geq J(\rho )\\
\infty &,\ \mbox{for }j<J(\rho )
\end{array} \right.\ ,
\label{mainres1}
\end{equation}
with $E_{tw} (j)$ from \eqref{etwip}.

\begin{figure}[t]
	\centering
	\mbox{\includegraphics[width=0.48\textwidth]{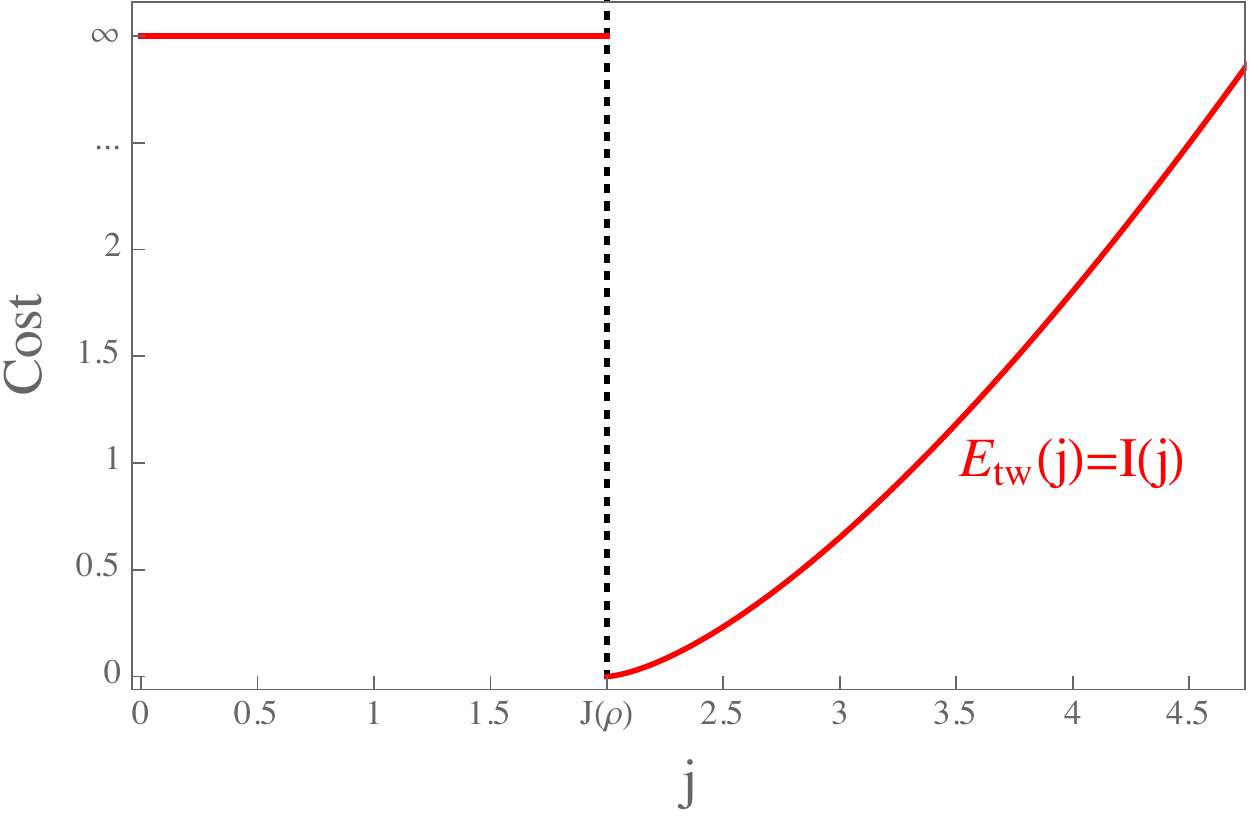}\quad\includegraphics[width=0.48\textwidth]{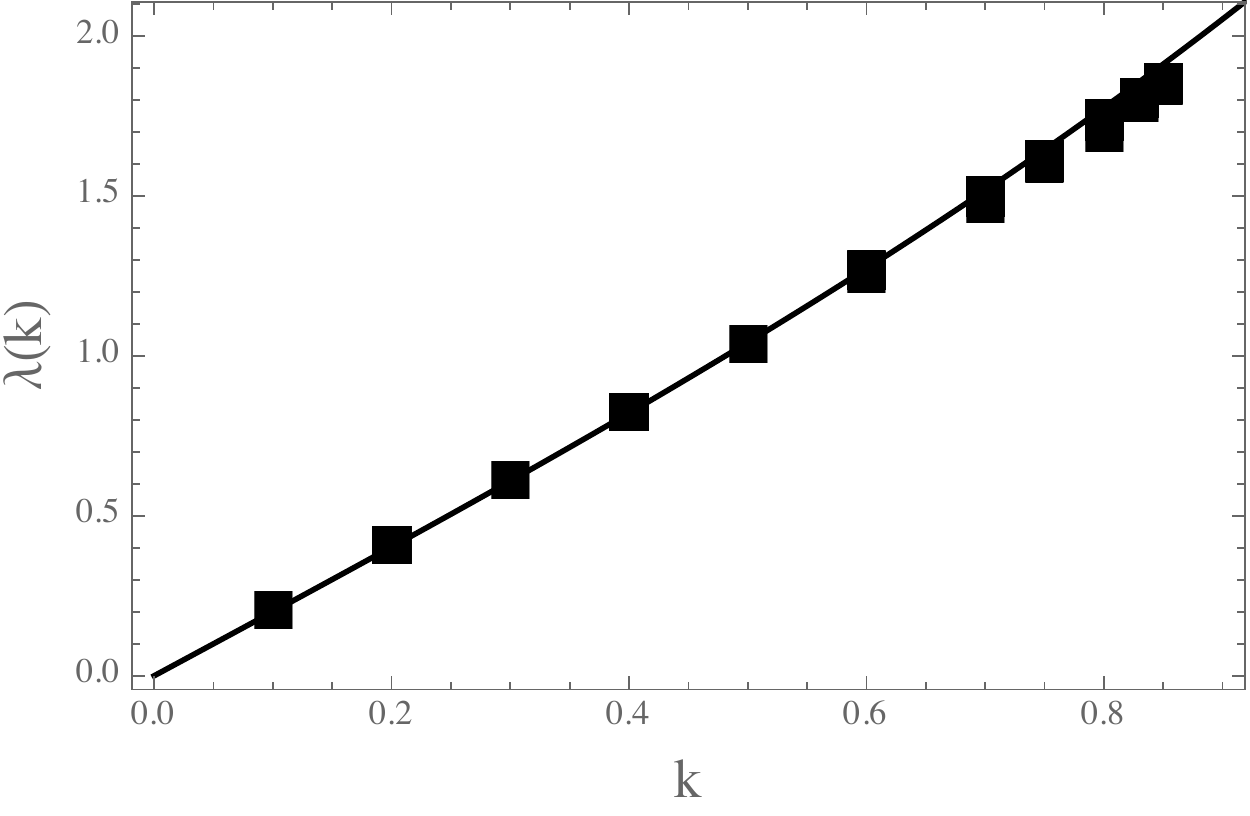}}
	\caption{For an IP with $\rho=1$ and $d=1$ we plot the rate function $I(j)$ given in \eqref{mainres1} on the left, and the corresponding SCGF on the right. Black diamond data points are obtained from a cloning algorithm simulation \cite{Giardina2011} with system size $L=128$, running time $L$ and $2^{15}$ clones. Error bars are of the size of the symbols.}\label{fig:costandSCGFip}
\end{figure}

\subsection{Predictions for lower current deviations\label{sec:iplow}}

As discussed in detail in \cite{pizzo} for ZRPs, a generic candidate for a phase separated profile to realize lower current deviations, irrespective of convexity of the flux function, is a condensed profile. 
This still holds for the IP, if a macroscopic number of particles concentrates on a fixed single lattice site we observe a lower bulk density of the system, and therefore the empirical current is also lower. 
To achieve a current $j<J(\rho)$ we require a bulk density
\[
R(j)= \frac12\big(\sqrt{d^2 +4j}-d\big)<\rho\ ,
\]
simply given by the inverse of $J(\rho)$ in \eqref{current_ip}, which is strictly increasing in $j$. Therefore the occupation number of the condensate is of order $L(\rho -R(j))$ and in order to stabilize it, the exit process from the condensate site has to be unusually slow. Its typical rate is of order $L(\rho -R(j))(d+\rho)$, and the empirically observed rate should be equal to the conditional current $j$. As explained in detail in \cite{pizzo}, the associated cost $E_c^L\left(j\right)$ to leading order in $L$ is given by the standard formula to slow down a Poisson process from the typical exit rate to the target rate $j$, which is given by
\begin{equation}
E_c^L\left(j\right)=L\left(\rho-R\left(j\right)\right)\left(d+R\left(j\right)\right)-j+j\ln\frac{j}{L\left(\rho-R\left(j\right)\right)\left(d+R\left(j\right)\right)} \ .
\label{ecip}
\end{equation}
Therefore, we have for diverging system size
\begin{alignat}{2}
e_{c} (j)\coloneqq\lim_{L\to\infty}\frac{E_c^L (j)}{L}= & \left(\rho-R\left(j\right)\right)\left(d+R\left(j\right)\right)\nonumber\\
=&J\left(\rho\right)-j-\rho\left(\rho-R\left(j\right)\right)\ ,
\label{ecj}
\end{alignat}
which is positive for all $j<J(\rho)$ and vanishes as $j\nearrow J(\rho)$. Note that the condensate can be interpreted as a boundary site with exit rate $j$, fixing the typical current in the bulk. Then the exit rate at the right end of the bulk into the condensate site is also increased. But since with $J'(\rho)>0$ all characteristic velocities are strictly positive, this only leads to a finite range boundary layer in the bulk density profile. This does not influence the overall current on a macroscopic scale, and therefore does not enter the cost in the rate function.

Another simple strategy to achieve a lower current deviation is to slow down the current across all bonds, independently, from $J(\rho)$ to $j<J(\rho)$. Again this leads to a cost $E_i^L$ proportional to $L$, which to leading order is given by 
\begin{equation}
e_i (j)\coloneqq\lim_{L\to\infty}\frac{E_i^L (j)}{L}=J\left(\rho\right)-j+j\ln\frac{j}{J\left(\rho\right)}.
\label{eij}
\end{equation}

For IPs the jump rates across a bond depend on both adjacent occupation numbers, suggesting hyperuniform states with alternating density profiles as further candidates to realize current large deviations (in analogy to results for exclusion \cite{popkov2010asep}). The simplest candidate is a profile alternating between densities $\rho_1\leqslant\rho\leqslant\rho_2$ with  $\rho_{1}+\rho_{2}=2\rho$. This corresponds to two typical currents across even and odd bonds, $\rho_{1}\left(d+\rho_{2}\right)<\rho_{2}\left(d+\rho_{1}\right)$. To stabilize this, we need to slow down the higher of both currents to the lower one. Eliminating one variable via $\rho_1=2\rho-\rho_2$, the target  current for this profile can be written as
\begin{equation}\label{eq:dunnoname}
j=\frac12\big( \rho_{1}\left(d+\rho_{2}\right)+\rho_{2}\left(d+\rho_{1}\right)\big) =\left(2\rho-\rho_{2}\right)\left(d+\rho_{2}\right)\ .
\end{equation}
The resulting cost per site, in the limit $L\to\infty$, is given by
\begin{alignat}{2}\label{eq:againdunno}
e_a^L\left(j\right) =&\frac{1}{2}\left(\rho_{2}\left(d+\rho_{1}\right)-j+j\ln\frac{j}{\rho_{2}\left(d+\rho_{1}\right)}\right)\nonumber\\
=&\frac{1}{2}\left(\rho_{2}(j)\left(d+2\rho-\rho_{2}(j)\right)-j+j\ln\frac{j}{\rho_{2}(j)\left(d+2\rho-\rho_{2}(j)\right)}\right)\ ,
\end{alignat}
where the prefactor $1/2$ takes into account that only half of the bonds are slowed down and
\eqref{eq:dunnoname} uniquely determines $\rho_2\left(j\right) >\rho$. 
We can check that these profiles in fact gives rise to currents $j<J(\rho)$ by differentiating \eqref{eq:dunnoname},
\begin{equation}
\frac{dj}{d\rho_{2}}=-d+2\rho-2\rho_{2}<0\quad\text{for all }\rho_{2}\geqslant\rho\ .
\end{equation}

Optimizing over all possible strategies, our prediction for the rate function per volume for lower current deviations $j<J(\rho)$ is given by the lower convex hull
\begin{equation}\label{eq:costfctlowerIP}
\begin{array}{cc}
\iota\left(j\right)\coloneqq\underline{\mbox{conv}}\{e_{a}(j),e_{c}(j),e_{i}(j)\} & \text{as }L\to\infty\end{array},
\end{equation}
as illustrated in Figure \ref{fig:transIP}. Our ansatz for the alternating profiles assumes product measures with alternating densities. In general, other hyperuniform states with long range correlations might have a different cost, but still proportional to the system size. 
By monotonicity arguments, one can see that other splits of the mass in alternating profiles should therefore not contribute to the rate function. 
They lead to costs in between $e_a\left(j\right)$ and $e_c\left(j\right)$, which can be seen as the two most extreme versions of splitting the mass. 
We see in Figure \ref{fig:transIP}, for two different parameter values $d$, that in general the alternating cost $e_a\left(j\right)$ is always below the condensed cost $e_c\left(j\right)$, coinciding only for $j=0$ an $J(\rho)$. 
For $j$ only slightly below $J(\rho)$ the dominating strategy is to slow down all bonds, whereas for smaller values of $j$ the alternating profile becomes dominant, leading to a rate function given by the lower convex hull of both curves. So our heuristic considerations predict a dynamic phase transition for lower current large deviations in IPs for large system sizes. Since the rate functions are of order $L$ and the rates of the IP are unbounded, these predictions are currently beyond verification with numerical methods which have been used to produce the data in Figure \ref{fig:costandSCGFip}.

\begin{figure}
	\centering
	\mbox{\includegraphics[width=0.48\textwidth]{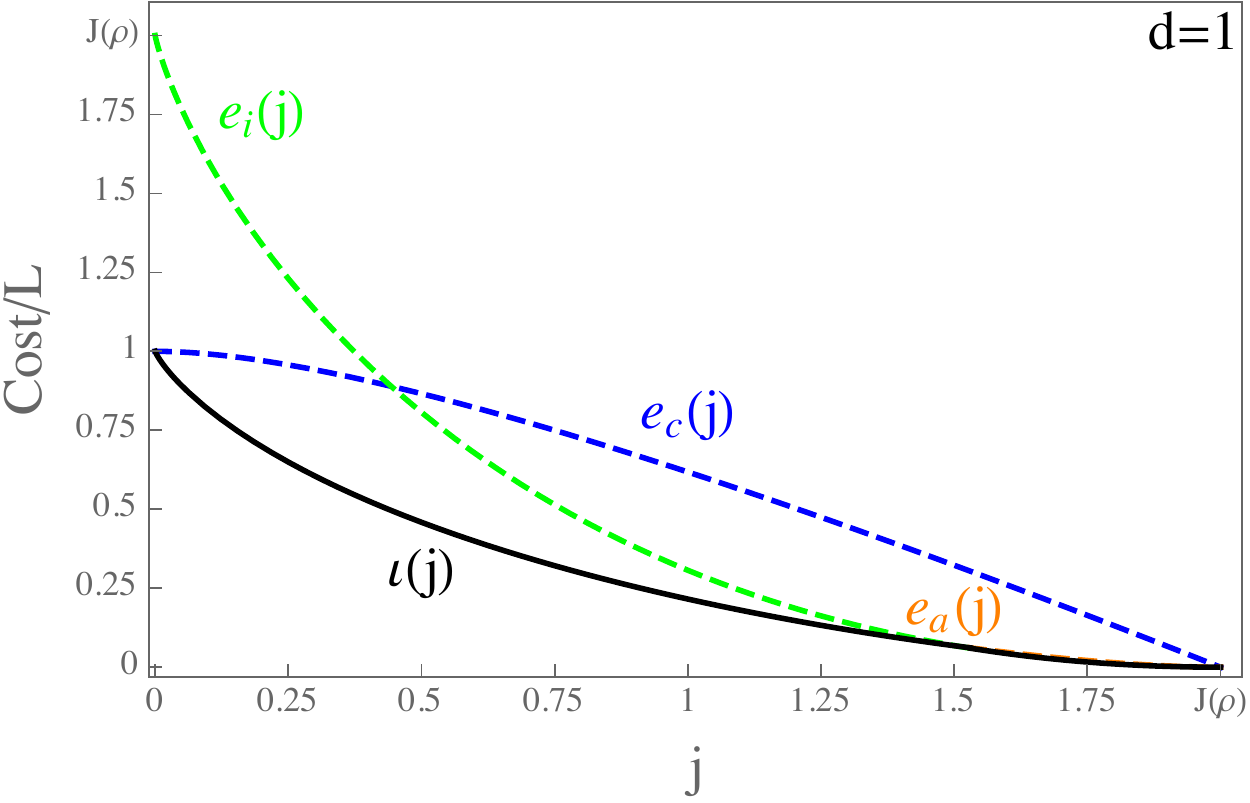}\quad\includegraphics[width=0.48\textwidth]{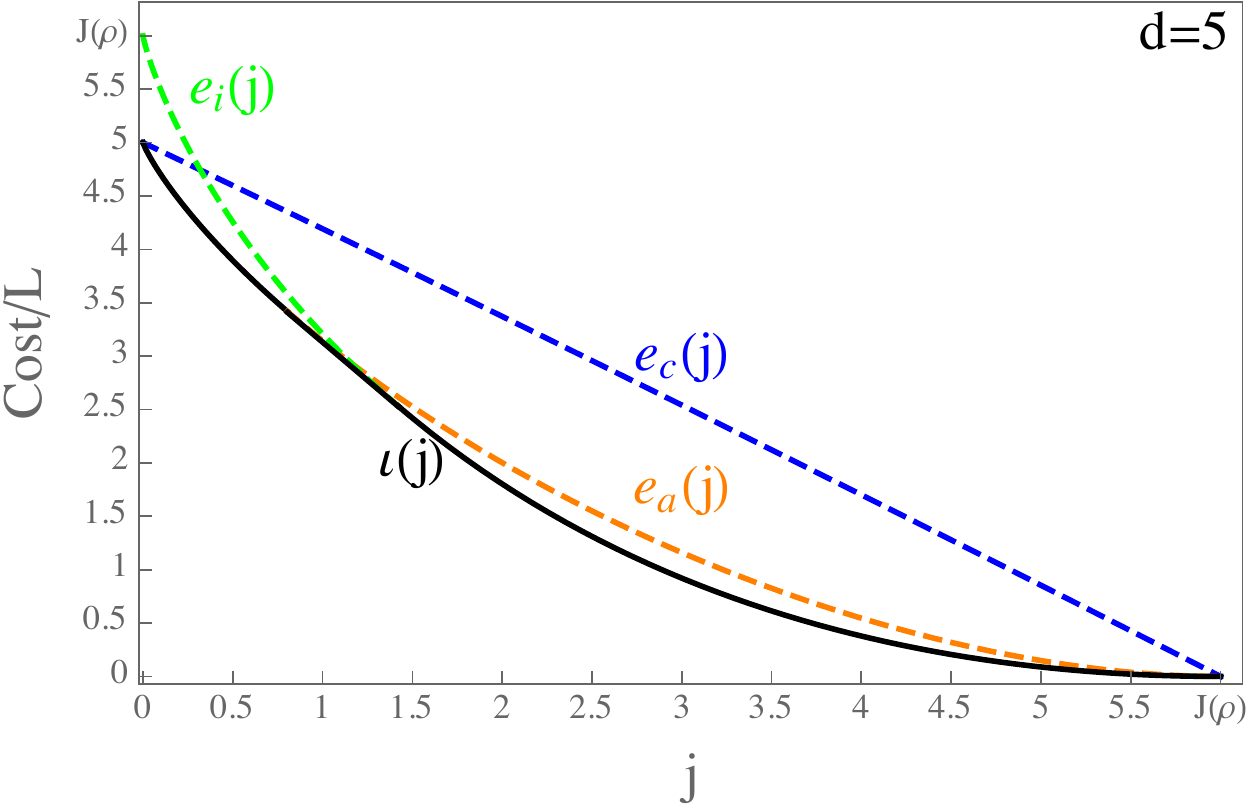}}
	\caption{Rate function for lower current deviations in the IP with $d=1$ (left) and $d=5$ (right). Plots show the intensive cost functions $e_c (j)$ \eqref{ecj}, $e_i (j)$ \eqref{eij}, $e_a (j)$ \eqref{eq:againdunno} and the corresponding rate function $\iota (j)$, given by the lower convex hull in \eqref{eq:costfctlowerIP}. The rate function is dominated by $e_i (j)$ close to $J(\rho)$ and by $e_a (j)$ for smaller values of $j$. As opposed to ZRPs, condensed states do not contribute to the large deviation.}\label{fig:transIP}
\end{figure}

%
%
\section{Partially asymmetric systems\label{sec:PA}}

We have demonstrated in the previous section and in \cite{pizzo} that the Jensen Varadhan approach for current large deviations, originally developed for the TASEP \cite{DerriBodJV}, can be applied more generally to totally asymmetric stochastic particle systems with convex or concave current-density relations. In this section we will investigate how it can be adapted to partially asymmetric systems. After describing the general approach and its relation to totally asymmetric systems, we will illustrate it for condensing ZRPs, which allow for the most interesting behaviour in this class of models, due to the interplay between travelling wave and condensed profiles.

\subsection{Travelling wave and condensed profiles}

To set the notation, we recall the general expression of stationary currents for systems with product measures in Section \ref{sec:2} and note that for partial and total asymmetry they are simply related as
\begin{equation}
J^{PA}\left(\rho\right) =\left(p-q\right)J^{TA}\left(\rho\right)\ ,
\end{equation}
where we use the labels $PA$ and $TA$ in the superscript to distinguish systems with $p(x,y)=p\delta_{y,x+1} +q\delta_{y,x-1}$ and $p(x,y) = \delta_{y,x+1}$, respectively.
We notice that the stationary measure \eqref{pm}, the relation $R(\phi )$ \eqref{density} and its inverse $\Phi (\rho )$ are unaffected by the partial asymmetry. Recalling $\eqref{eq:entrflux}$, this implies that the thermodynamic entropy $h(\rho)$ remains unchanged. 
Using \eqref{eflux} we can determine the corresponding entropy flux $g(\rho )$ and get
\begin{equation}
h^{PA}\left(\rho\right)=h^{TA}\left(\rho\right)\quad\mbox{and}\quad g^{PA}\left(\rho\right)=\left(p-q\right)g^{TA}\left(\rho\right)\ .
\end{equation}
From \eqref{entprod}, the Jensen Varadhan functional for partially asymmetric systems is then simply given by
\begin{equation}
\mathcal{F}^{PA}\left(\rho_{1},\rho_{2}\right)=\left(p-q\right)\mathcal{F}^{TA}\left(\rho_{1},\rho_{2}\right).
\end{equation}
Again, in the same way, the consistency condition for travelling wave profiles \eqref{cond} becomes
\begin{alignat}{2}
G^{PA}\left(\rho_{1},\rho_{2}\right)=&\frac{\rho\left(J^{PA}\left(\rho_{2}\right)-J^{PA}\left(\rho_{1}\right)\right)-J^{PA}\left(\rho_{2}\right)\rho_{1}+J^{PA}\left(\rho_{1}\right)\rho_{2}}{\rho_{2}-\rho_{1}}\\
=&\left(p-q\right)G^{TA}\left(\rho_{1},\rho_{2}\right)=j.
\end{alignat}
The optimization expression for the cost function \eqref{twcost} can then be used to determine its partially asymmetric counterpart,
\begin{alignat}{2}
E_{tw}^{PA}\left(j\right)=&\inf_{\rho_1 \leq\rho\leq\rho_2}\Big\{ \mathcal{F}^{PA}\left(\rho_{1},\rho_{2}\right):G^{PA}\left(\rho_{1},\rho_{2}\right)=j\Big\}\nonumber\\
=&(p-q)\inf_{\rho_1 \leq\rho\leq\rho_2}\Big\{ \mathcal{F}^{TA}\left(\rho_{1},\rho_{2}\right):G^{TA}\left(\rho_{1},\rho_{2}\right)=j/\left(p-q\right)\Big\}\nonumber\\
=&(p-q)E_{tw}^{TA}\left( \frac{j}{p-q}\right)\ .
\label{eq:PAtravellingwaveCOST}
\end{alignat}

In analogy to \eqref{ecip}, recalling that for partial asymmetry the bias across each bond is multiplied by $(p-q)$, we can write the cost of a condensed profile as
\begin{equation}
E_{c}^{PA}\left(j\right)=\left(p-q\right)u\left(\left(\rho-R\left(j\right)\right)L\right)-j+j\ln\frac{j}{\left(p-q\right)u\left(\left(\rho-R\left(j\right)\right)L\right)}\ .
\end{equation}
Pulling out the factor $(p-q)$ it is easy to see that the same scaling relationship as for $E_{tw}^{PA}$ holds in general for condensed states, i.e.
\begin{align}\label{eq:epasonoesausto}
E_{c}^{PA}\left(j\right)=(p-q)E_{c}^{TA}\left(\frac{j}{p-q}\right)\ .
\end{align}
We will use this general understanding of the effects of partial asymmetry on costs of travelling wave and condensed profiles in the next subsection and apply it to condensing ZRPs.

Note that in the above derivation we do not make use of the factorized form of stationary states. The only important ingredient is that states, and their associated entropy, are independent of the bias $(p,q)$. Then the above relation between costs for partially and totally asymmetric systems holds.

\subsection{Partially asymmetric ZRPs\label{sec:zrp}}

\begin{figure}[t]
	\centering
	\includegraphics[scale=0.75]{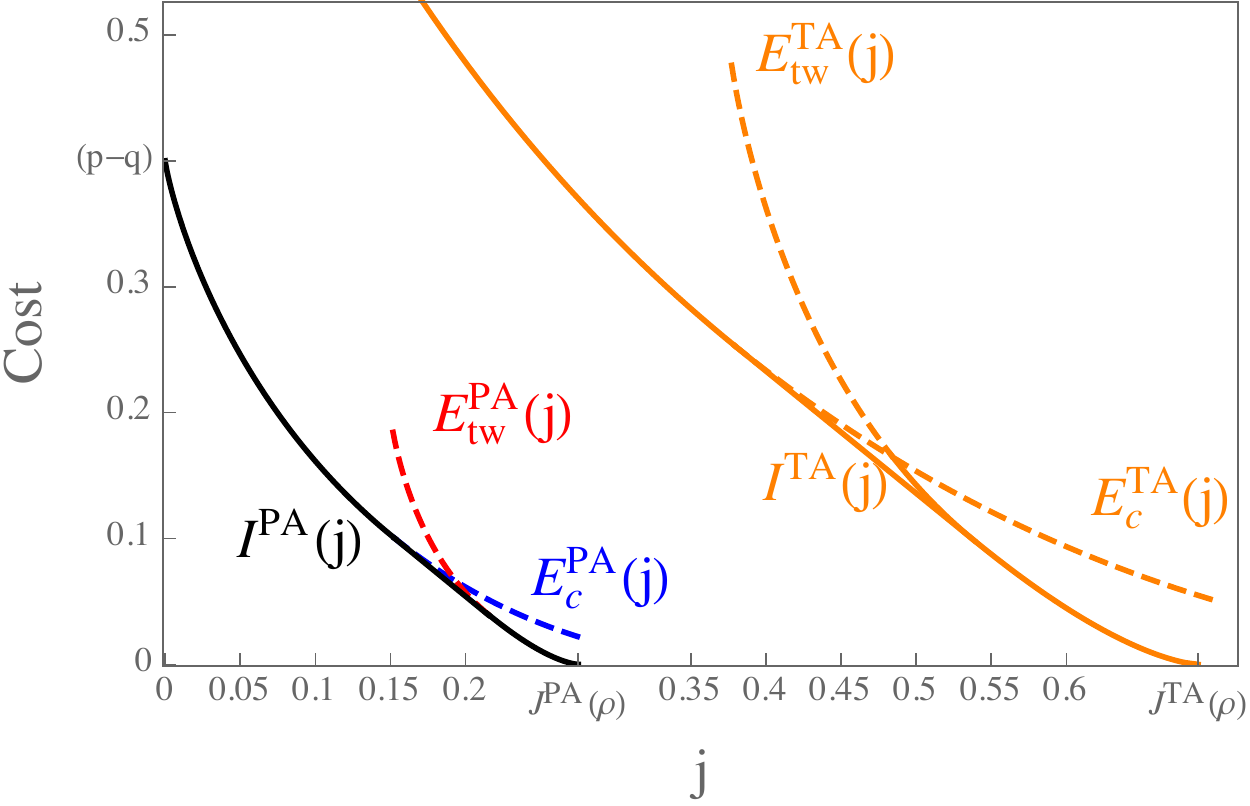} 
	\caption{Large deviation rate function (black) for a partially asymmetric ZRP with rates \eqref{conrates} where $b=3.5$, $\rho=0.25$ and $p=0.7$, resulting from the convex hull \eqref{eq:convexhullPAzrp}. (Red) JV cost function \eqref{eq:PAtravellingwaveCOST}. (Blue) Condensate cost function \eqref{eq:epasonoesausto}. (Orange) Cost functions for totally asymmetric dynamics, see \cite{pizzo} for details including also the limited range of $E_{tw} (j)$. Rate functions $I^{PA}(j)$ and $I^{TA} (j)$ are related by a scaling given in \eqref{eq:PAtravellingwaveCOST}.}\label{fig:costJVPAzrpcond}
\end{figure}
\begin{figure}[t]
	\centering
	\includegraphics[scale=0.75]{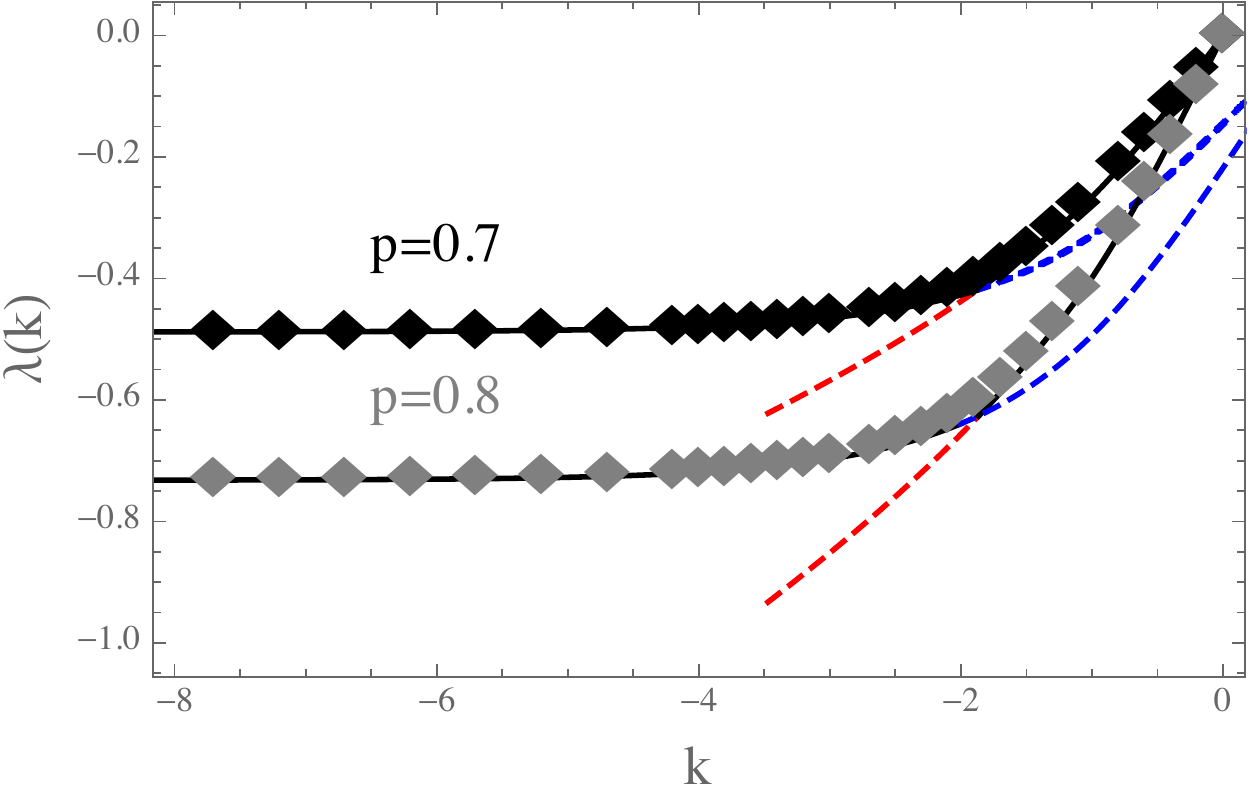} 
	\caption{SCGF (black line) given by \eqref{eq:momgen} as $L\to\infty$ for a partially asymmetric ZRP with rates \eqref{conrates} where $b=3.5$, $\rho=0.25$ and $p=0.7$ and $0.8$, resulting from Legendre transform of the rate function \eqref{eq:convexhullPAzrp}. 
Data points obtained from a simulation using the cloning algorithm (with $2^{15}$ clones, $L=64$ and running time $L^2$). The red (blue) dashed curve is the Legendre transform of the travelling wave (condensed) cost functions $E_{tw}^{PA} (j)$ and $E_{c}^{PA} (j)$. The resulting kink corresponds to the dynamic transition between travelling wave and condensed profiles. Error bars are of the size of the symbols.}\label{fig:SCGFzrpdiffp}
\end{figure}

We will illustrate the general result of the previous section for lower current deviations for a condensing ZRP with rates
\begin{equation}\label{conrates}
u(0)=0\ ,\quad u(n)=1+b/n\quad\mbox{for all }n\geq 1\ .
\end{equation}
This model has been widely studied in the literature (see e.g.\ \cite{evansBrazil,giardina2010correlation,grosskinsky2011condensation}).
It is known to have a concave flux function, and in \cite{pizzo} we derived the current large deviation function for the totally asymmetric version of the process. It exhibits a dynamic phase transition, where for certain parameter values the rate function is determined by condensed states rather than travelling waves for small enough $j$, as is shown in Figure \ref{fig:costJVPAzrpcond}.
For the rates \eqref{conrates} we have $u\left(\left(\rho-R\left(j\right)\right)L\right)\simeq 1$, as $L\to\infty$,  to leading order, and therefore the condensed cost \eqref{eq:epasonoesausto} simplifies to
\begin{equation}\label{eq:epasonoesausto2}
E_{c}^{PA}\left(j\right)\simeq\left(p-q\right) \left(1-\frac{j}{p-q} +\frac{j}{p-q}\ln \frac{j}{p-q}\right)\ .
\end{equation}
This asymptotic behaviour is independent of $L$ since the rates of the process are bounded, in contrast to the IP. 
The main result in \cite{pizzo} states that the rate function for lower current deviations in any ZRP with concave flux function is the lower convex hull of $E_{tw}^{TA}$ and $E_c^{TA}$. With \eqref{eq:PAtravellingwaveCOST} and \eqref{eq:epasonoesausto} the same is true for partially asymmetric systems, i.e.
\begin{equation}\label{eq:convexhullPAzrp}
\begin{array}{ccc}
I^{PA}\left(j\right)=\mathrm{\underline{conv}}\left\{ E_{tw}^{PA},E_{c}^{PA}\right\} \left(j\right) & \text{for all }0<j<J\left(\rho\right), & \text{if }\rho_{c}<\infty\end{array}
\end{equation}
for condensing systems, and 
\begin{equation}\label{eq:convexhullPAzrp2}
\begin{array}{ccc}
I^{PA}\left(j\right)= E_{tw}^{PA} (j) & \text{for all }0<j<J\left(\rho\right), & \text{if }\rho_{c}=\infty\end{array}
\end{equation}
for non-condensing systems. 
In particular, we can relate the full rate function, $I^{PA}$, to the totally asymmetric one for general ZRPs, using the scaling derived above, as
\begin{align}
\label{eq:IPA}
  I^{PA}\left(j\right) =(p-q)\, I^{TA}\left(\frac{j}{p-q}\right)\ ,
\end{align}
which is illustrated in Figure \ref{fig:costJVPAzrpcond} for the particular example with rates \eqref{conrates}. The travelling wave cost for $p=1-q=1$ has been evaluated numerically for the given parameter values in \cite{pizzo}, in analogy to the procedure outlined in Section \ref{sec:inclusion} for the IP. 
We do not discuss details of the shape of the cost and rate function here, which can be found in \cite{pizzo}. 
In Figure \ref{fig:SCGFzrpdiffp} we numerically confirm this result by comparing simulation data to the SCGF $\lambda (k)$, predicted as the Legendre transform of the rate function $I^{PA} (j)$.

\subsection{Beyond phase separated states}

In general, outside the accessible range of conditional currents for travelling wave or condensed profiles (which is $0<j<J(\rho)$ for ZRPs with concave flux function), we expect the cost for current large deviations to scale with the system size as explained in Section \ref{sec:iplow} for the IP. 
The presence of partial asymmetry introduces additional randomness and allows fluctuations also at the level of the spatial part of the dynamics. In fact, it is possible to reach a target current $j$, conditioning on an empirical bias $(p',q')$ with
\begin{equation}
\begin{array}{cc}
\left(p'-q'\right)\Phi\left(\rho\right)=j & \text{and}\quad p'+q'=1\end{array},
\end{equation}
which implies
\begin{equation}
p'\left(j\right)=\frac{j+\Phi\left(\rho\right)}{2\Phi\left(\rho\right)}.
\end{equation}
Due to the obvious constraints $p',q'\in [0,1]$ on the spatial coefficients, it is possible to achieve a bounded set of currents $j\in\left[-J^{TA}\left(\rho\right),J^{TA}\left(\rho\right)\right]$ by conditioning purely on an empirical bias, and leaving the jump rates in the process unchanged. In particular, due to the additional randomness from partial asymmetry, the system may obtain atypical fluctuations of the current in the direction opposite to the stationary current. The cost to alter the empirical bias across the whole system is of order $L$ (since it independently accrues at each bond) and diverges with the system size. The cost to condition on an atypical spatial bias per bond is given by the relative entropy, which is the standard rate function for observing an empirical bias $\left(p',q'\right)$, given the true bias $\left(p,q\right)$ (see e.g.\ \cite{Hollander}), which leads to
\begin{equation}\label{eq:costbwcur}
e_{pq}\left(j\right)
\coloneqq \Phi\left(\rho\right)\left(p'\left(j\right)\ln\frac{p'\left(j\right)}{p}+q'\left(j\right)\ln\frac{q'\left(j\right)}{q}\right).
\end{equation}

To complete the picture, it is possible to achieve currents beyond the interval $\left[-J^{TA}\left(\rho\right),J^{TA}\left(\rho\right)\right]$ by increasing also the empirical jump rates of the model. In fact, using the same reasoning as for the condensed case in \eqref{eq:epasonoesausto2},  increasing the empirical exit rate of the particles $\Phi\left(\rho\right)$, has a cost function per site given by the acceleration of a Poisson process from $\Phi\left(\rho\right)$ to a value $\hat{\phi}>\Phi\left(\rho\right)$
\begin{equation}
e_{\Phi}\left(\hat{\phi}\right)
\coloneqq
\Phi\left(\rho\right)-\hat{\phi}+\hat{\phi}\ln\frac{\hat{\phi}}{\Phi\left(\rho\right)}.
\end{equation}
Keeping the asymmetry $\left(p,q\right)$ fixed, this mechanism in isolation 
with $j=\left(p-q\right)\hat{\phi}$ would lead to a cost function per volume
\begin{equation}\label{eq:costphimorto}
e_{J}\left(j\right)\coloneqq\left(p-q\right)e_{\Phi}\left(\hat{\phi}\right)=j-J^{PA}\left(\rho\right)+J^{PA}\left(\rho\right)\ln\frac{J^{PA}\left(\rho\right)}{j},
\end{equation}
where $J^{PA}\left(\rho\right)=\left(p-q\right)\Phi\left(\rho\right)$. In general the two mechanisms \eqref{eq:costbwcur} and \eqref{eq:costphimorto} can interact. For instance, to obtain an atypical negative current, the system needs to change the spatial bias, but it may be more efficient to combine this with an additional increase of the empirical exit rate. Along the same lines, reaching a current above $J^{PA}\left(\rho\right)$ can be achieved as a combination of increasing the asymmetry of the spatial part and the system activity. Adding both costs with 
\begin{equation}
\left(p'-q'\right)\hat{\phi}=j\quad\mbox{and thus}\quad p'\left(j,\hat{\phi}\right)=\frac{j+\hat{\phi}}{2\hat{\phi}}\ ,
\end{equation}
leaves only $\hat\phi$ as a free parameter to optimize. 
This leads to the combined optimal cost
\begin{alignat}{2}\label{eq:costoptspac}
e_{pq;J}\left(j\right)\coloneqq\min_{\hat{\phi}\geq\Phi (\rho)}\Bigg\{&\hat{\phi}\left(p'\left(j,\hat{\phi}\right)\ln\frac{p'\left(j,\hat{\phi}\right)}{p}+q'\left(j,\hat{\phi}\right)\ln\frac{q'\left(j,\hat{\phi}\right)}{q}\right)\nonumber\\
&+\left(p-q\right)\left(\Phi\left(\rho\right)-\hat{\phi}+\hat{\phi}\ln\frac{\hat{\phi}}{\Phi\left(\rho\right)}\right)\Bigg\}.
\end{alignat}
\begin{figure}[t]
	\centering
	\includegraphics[scale=0.75]{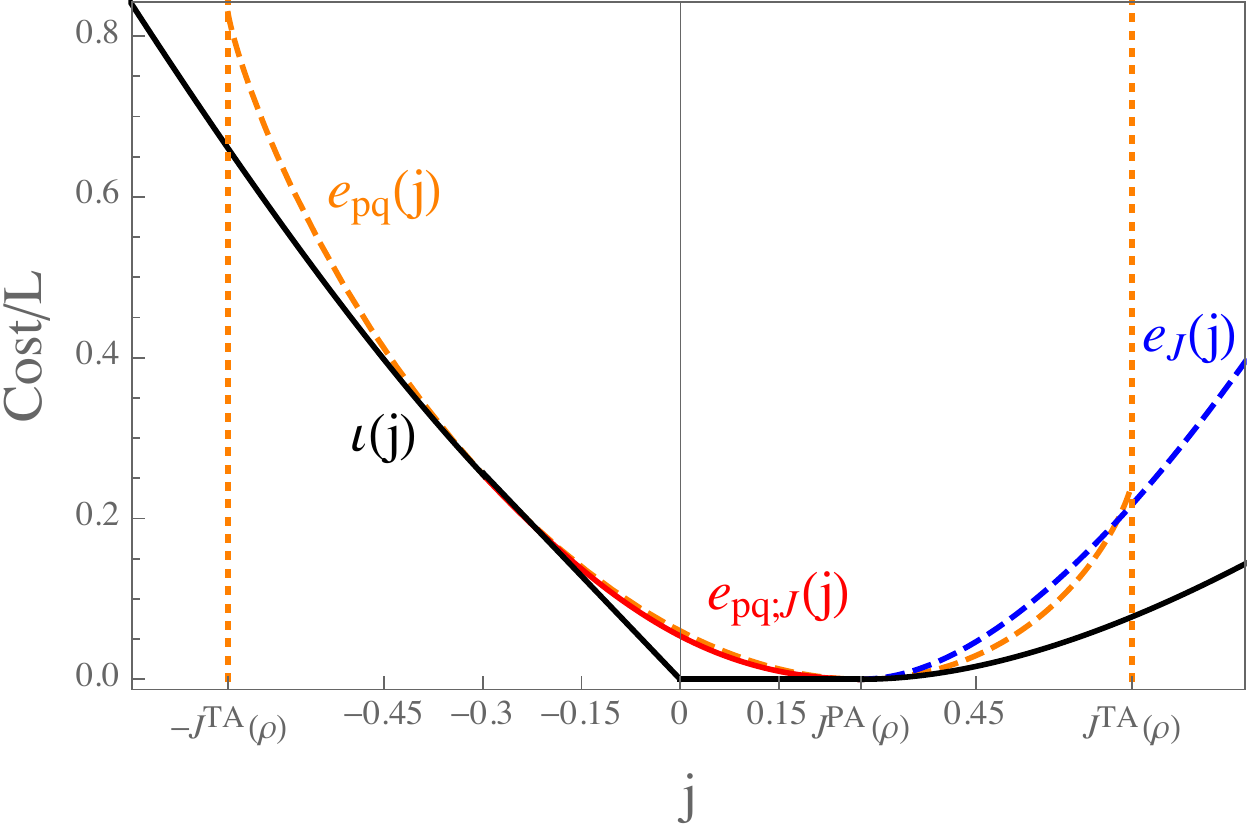} 
	\caption{Large deviation rate function $\iota (j)$ \eqref{eq:convexhullcontrib} per volume (black) for a partially asymmetric ZRP with rates \eqref{conrates} where $b=3.5$, $\rho =0.25$ and $p=0.7$.  	
	The cost of travelling wave profiles for $j \in [0,J^{PA}(\rho)]$ vanishes on the scale $L$, leading to a dynamic transition from phase separated profiles to global activity and bias conditioning, as explained in the text. These qualitative features of the plot are independent of the choice of jump rates (which would only shift the position of $J^{PA}$). The cost $e_{pq;J} (j)$ optimizing between spatial and activity contribution given in \eqref{eq:costoptspac} (full red line) does not coincide with $e_{pq} (j)$ \eqref{eq:costbwcur} (dashed orange) nor $e_J (j)$ \eqref{eq:costphimorto} (dashed blue).
}\label{fig:COSTpaZRP}
\end{figure}

Finally, we denote the cost per volume of the $L$-independent rate function \eqref{eq:IPA} due to phase-separated profiles as
\begin{equation}
\iota^{PA}\left(j\right)=\lim_{L\to\infty}\frac{I^{PA}\left(j\right)}{L}=\left\{ \begin{array}{cl}
0 &,\ j\in\left[0,J^{PA}\left(\rho\right)\right]\\
\infty &,\ \text{otherwise}
\end{array}\right. .
\end{equation} 
Then our prediction for the rate function per volume for any $j\in\mathbb{R}$ can be written as the convex hull
\begin{equation}\label{eq:convexhullcontrib}
\begin{array}{cc}
\iota\left(j\right)\coloneqq\mathrm{\underline{conv}}\left\{ e_{pq;J},\iota^{PA}\right\} \left(j\right) & \mbox{for all } j\in\mathbb{R}.
\end{array}
\end{equation}
The plot of all relevant cost functions and the resulting rate function can be found in Figure \ref{fig:COSTpaZRP}. For small and large values of $j$ the rate function is dominated by the combined cost $e_{pq;J} (j)$, but vanishes for the whole interval $j\in [0,J^{PA} (\rho )]$, due to the size-independent cost of phase separated profiles described in the previous subsection. This leads to a dynamic transition for negative $j$ close to the origin, corresponding to a mixture of fully condensed profiles with vanishing current and homogeneous ones with global change of activity/bias and negative current. 
We obtain the corresponding SCGF by Legendre-Fenchel transform of \eqref{eq:convexhullcontrib}, which is shown in Figure \ref{fig:SCGFpaZRP} in comparison with simulation data. 
This is possible here, in contrast to results in Section \ref{sec:iplow} for IPs, since the rates of the ZRP we consider are bounded. The two affine parts of the rate function turn into kinks, while the kink at $j=0$ turns into the flat part of the SCGF. Since the large deviation speed now scales with $L$, the SCGF \eqref{eq:momgen} has to be rescaled as
\begin{equation}\label{scgfi}
\frac{1}{L}\lambda^{L}\left(kL\right)\to\lambda_{\infty}\left(k\right)\quad\mbox{as }L\to\infty\  ,
\end{equation}
to compare data with tilt parameter $kL$ to the asymptotic behaviour.

\begin{figure}[t]
	\centering
	\includegraphics[scale=0.75]{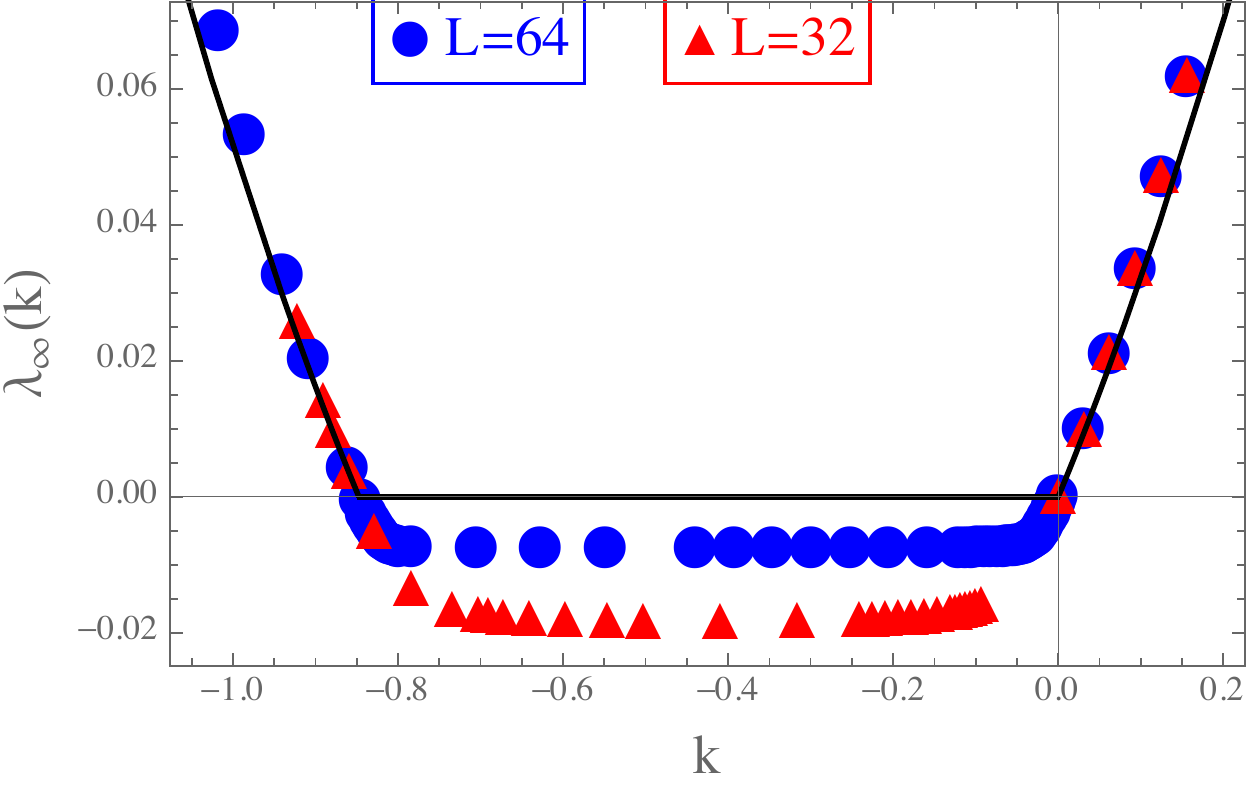} 
	\caption{SCGF $\lambda_\infty\left(k\right)$ normalized by the system size (black line) given by \eqref{scgfi},  corresponding to the rate function \eqref{eq:convexhullcontrib} for the same ZRP as in Figure \ref{fig:COSTpaZRP}. Data points are obtained from a simulation using the cloning algorithm with $2^{15}$ clones, $L^2$ running time and system size $L=32$ (red), $L=64$ (blue). The discrepancy between the data points and $\lambda_\infty\left(k\right)$ is due to a generic finite size effect smoothing kinks and affine parts of the function, which decreases with $L$. Note that the fluctuation relation \eqref{fdrel} holds as explained in the text with $V=\ln p/q\approx 0.847$. Error bars are of the size of the symbols.}\label{fig:SCGFpaZRP}
\end{figure}

Note that for ZRPs, alternating profiles as discussed in Section \ref{sec:inclusion} do not provide different currents $j$ than the phase separated ones, since jump rates depend only on the departure site occupation, and therefore such states do not have to be considered. 
Furthermore, in the large deviation rate function per volume in Figure \ref{fig:COSTpaZRP} the fine details for $j\in [0,J^{PA} (\rho )$ (shown in Figure \ref{fig:costJVPAzrpcond}) are scaled away. 
In particular, the interplay between condensed and travelling wave profiles in this region is irrelevant on that scale, and in fact we expect $\iota (j)$ to show the same qualitative features for any ZRP with concave flux-density relation. 
The costs entering the rate function \eqref{eq:convexhullcontrib} only depend on macroscopic features, and can in principle be computed for much more general models. 
However, as we have seen in Section \ref{sec:iplow} for the IP, microscopic features of the system can lead to other profiles contributing to the rate function, and \eqref{eq:convexhullcontrib} is not a completely general expression. 
Still, for models with finite correlation lengths in the stationary state and concave flux-density relation, we expect the rate function to show the same qualitative behaviour and in particular to exhibit a crossover between local phase separated profiles and global conditioning. Typical profiles contributing to data points in Figure \ref{fig:SCGFpaZRP} for different values of $k$ are shown in Figure \ref{fig:illu4}.

In Figure \ref{fig:SCGFpaZRP} it is clearly visible that the generalized fluctuation relation for the current \cite{Lebowitz1998,harris75breakdown,rakos2008range} is obeyed by the system, which translates into a corresponding symmetry property of the SCGF
\begin{equation}\label{fdrel}
\lambda^L (k) =\lambda^L \big(-k-VL\big) \quad\mbox{for all }k\in\R\mbox{ and with }V\coloneqq\ln \frac{p}{q}\ .
\end{equation}
As explained in \cite{rakos2008range}, the parameter $V/2$ can be interpreted as the field per volume driving the original system, which has to be reversed to achieve a negative current deviation. Note that $V$ and therefore the symmetry \eqref{fdrel} is entirely independent of the rates $u(n)$ of the process.

\begin{figure}[t!]
\begin{center}
\mbox{\includegraphics[width=0.48\textwidth]{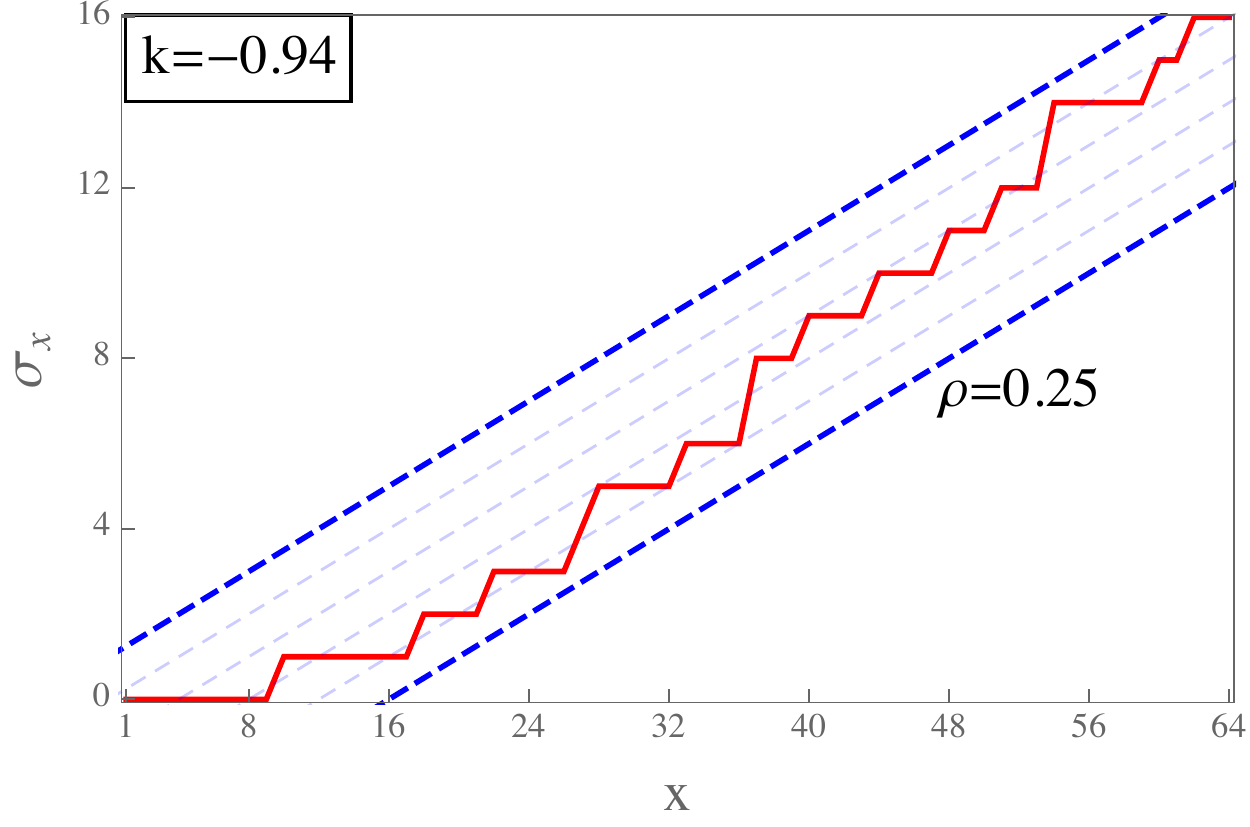}\quad\includegraphics[width=0.48\textwidth]{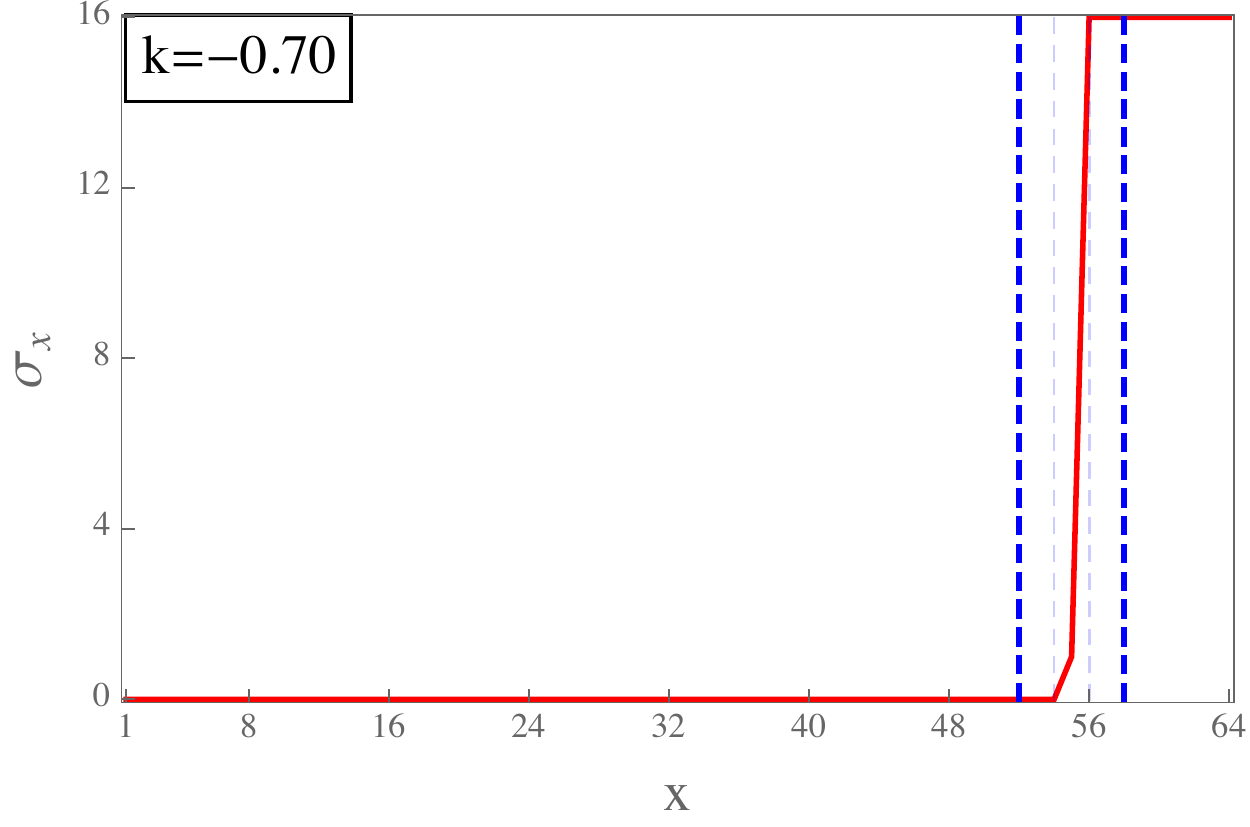}}\\
\mbox{\includegraphics[width=0.48\textwidth]{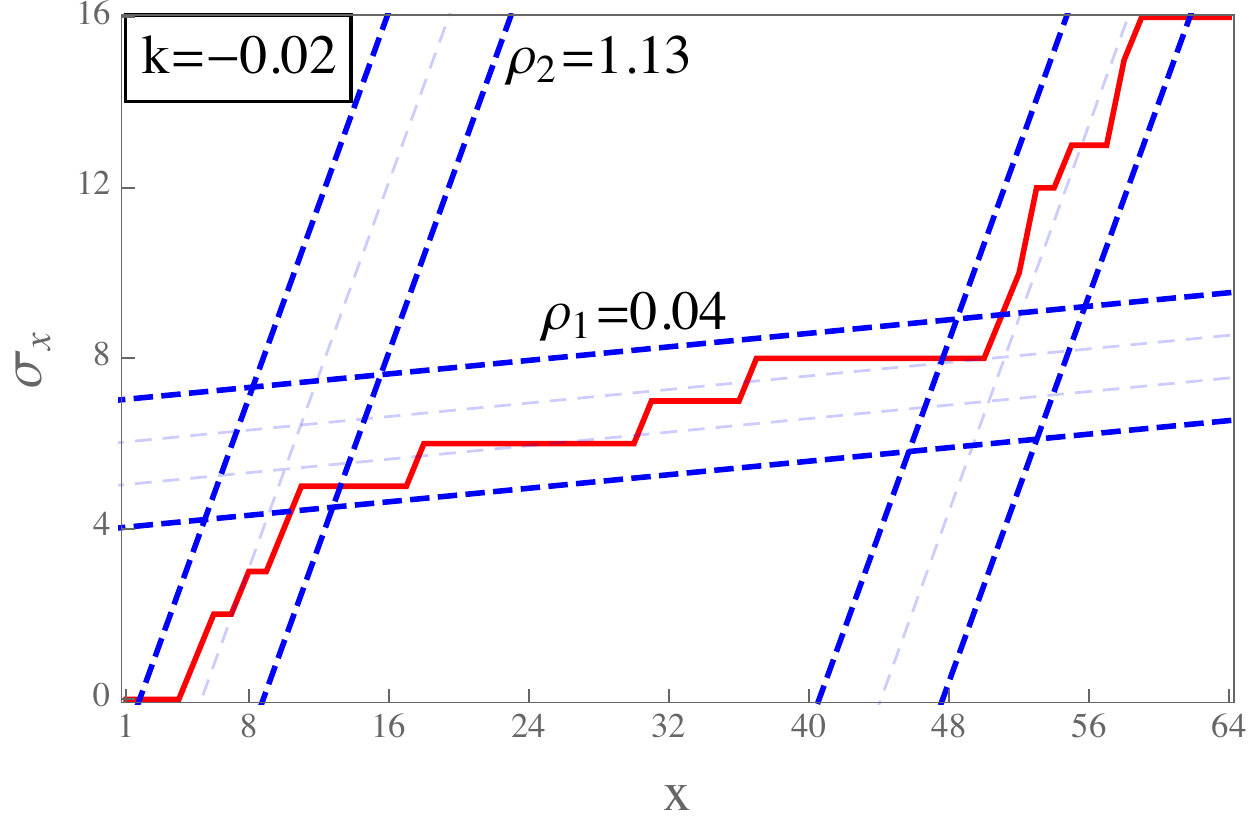}\quad\includegraphics[width=0.48\textwidth]{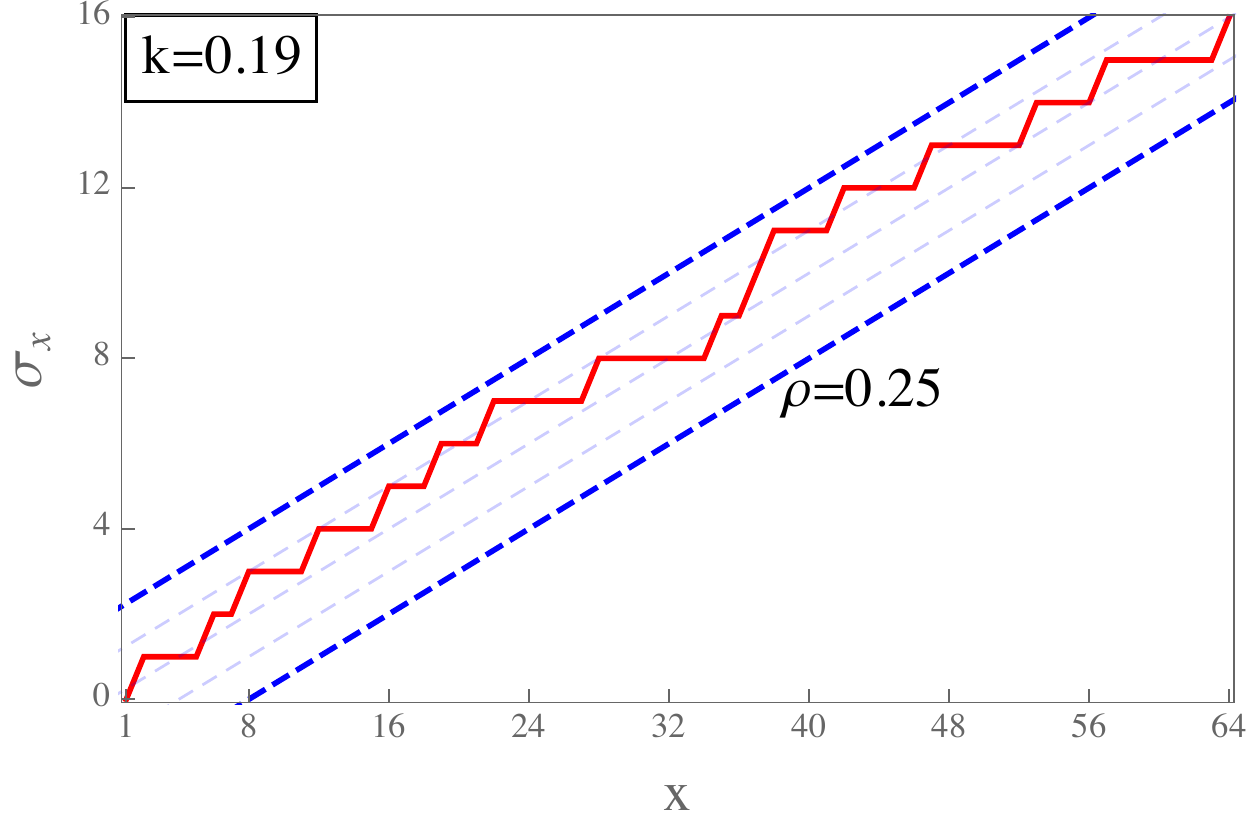}}
\end{center}
\caption{\label{fig:illu4}
Typical integrated profiles $\sigma_x \coloneqq\sum_{y\leq x} \eta_y$ (red full lines) that contribute to data points in Figure \ref{fig:SCGFpaZRP} for different values of $k$, using the same ZRP as in Figure \ref{fig:COSTpaZRP}. Dashed blue lines illustrate the corresponding densities $\rho =0.25$ for flat profiles with $k=-0.94$ and $0.19$, the optimal pair $(\rho_1 ,\rho_2 )$ for travelling wave profiles for $k=-0.02$ minimizing \eqref{twcost}, and the condensate for $k=-0.7$. 
}
\end{figure}

%
%
%
%
\section{Discussion\label{sec:discussion}}

We explore the general applicability of a recent approach \cite{pizzo} to study current large deviations in periodic particle systems with unbounded local occupation number. We cover extensions to convex current-density relations and also partially asymmetric dynamics, which we illustrate for IPs and ZRPs. 
In addition to phase separated profiles, which lead to rate functions independent of the system size in a restricted range of currents, we also predict extensive rate functions outside this range. While the particular profiles contributing to the extensive costs depend on the particular model (as illustrated here for IPs and ZRPs), the qualitative features of the resulting extensive rate function (cf.\ Figure \ref{fig:COSTpaZRP}) are expected to be quite generic: vanishing for currents accessible by phase separated profiles, and an associated crossover to extensive costs.

We have also established a general relationship between partially and totally asymmetric costs for travelling wave and condensed profiles. This holds for any model where the stationary state (whether it factorizes or not) does not depend on the bias of the dynamics. 
Whenever the stationary state has finite correlation lengths travelling wave profiles provide size-independent costs to realize large deviations in a restricted range of currents, depending on the convexity properties of the current-density relation. The approach based on Jensen-Varadhan theory is expected to apply also for non-factorized steady states, and in this case the thermodynamic entropy would be characterized by the thermodynamic limit of canonical entropies
\[
\frac{1}{L}\log Z_{L,N} \to h(\rho )\quad\mbox{as }L,N\to\infty ,\ N/L\to\rho\ .
\]
Existence of this limit would be a minimal prerequisite to apply the approach at least on a heuristic level. 
In this context Katz-Lebowitz-Spohn models \cite{kls} which have explicitly known non-factorized states of nearest-neighbour Gibbs type provide a promising class to further test the applicability of this approach. The current-density relations of those systems are also known to exhibit convex and concave regions for certain parameter values, leading to interesting phase diagrams for open boundaries \cite{popkovschuetz}. While fully concave or convex current-density relations are covered in the present paper, the interplay between convex and concave regions is likely to lead to interesting effects for travelling wave profiles.

%
%
%
%
\section*{Acknowledgements}
AP acknowledges support by the Engineering and Physical Sciences Research Council (EPSRC), Grant No. EP/L505110/1, by The Alan Turing Institute EPSRC grant EP/N510129/1 and seed project SF029 ``Predictive graph analytics and propagation of information in networks"

%
%
%
%

%
%
%
%






\end{document}